\documentclass[11pt,preprintnumbers,nofootinbib,amsmath,amssymb]{revtex4}
\usepackage{float}
\usepackage{indentfirst}
\usepackage{amsmath,amssymb,epsfig,subfigure}
\usepackage{graphicx}
\usepackage[export]{adjustbox}
\usepackage{color}
\usepackage{multirow}
\usepackage{booktabs}
\usepackage{changes}
\usepackage{footnote}
\usepackage{mathrsfs}

\usepackage{hyperref}
\hypersetup{colorlinks=true,linkcolor=blue,citecolor=magenta}
\begin{document}
\pagestyle{plain}
\renewcommand{\baselinestretch}{1.15}
\makeatletter
\renewcommand{\thesubfigure}{\alph{subfigure}}
\renewcommand{\@thesubfigure}{(\thesubfigure)\hskip\subfiglabelskip}
\makeatother

\title{Boson Stars with Ayón-Beato-García nonlinear electrodynamics}

\preprint{}

\author{Long-Xing Huang$^{1,2}$ and Yong-Qiang Wang$^{1,2}$\footnote{Corresponding author. E-mail: yqwang@lzu.edu.cn}}

\affiliation{$^{1}$ Lanzhou Center for Theoretical Physics, Key Laboratory of Theoretical Physics of Gansu Province, School of Physical Science and Technology, Lanzhou University, Lanzhou 730000, China\\
 $^{2}$ Institute of Theoretical Physics $\&$ Research Center of Gravitation, Lanzhou University, Lanzhou 730000, China}


\date{\today}

\begin{abstract}
In this paper, we study spherically symmetric boson star solutions in Einstein gravity minimally coupled to Ayón-Beato–García nonlinear electrodynamics and a massive complex scalar field. We find that these solutions exist only when the electric charge $q$ is below a critical value. Within this range, solutions with relatively small charges can exhibit a gravitational blueshift, and the radial mass function becomes negative in some regions while the ADM mass remains positive. For larger charges, frozen boson-star solutions emerge as the scalar-field frequency $\omega$ approaches zero. In these configurations, the metric function $1/g_{rr}$ and $-g_{tt}$ become very small near a special position $r_{\mathrm{ch}}$, and the spacetime resembles an extremal black hole from the perspective of an observer at infinity. Moreover, timelike trajectories in these frozen boson stars are strongly deflected near $r_{\mathrm{ch}}$, and particles require a longer coordinate time to traverse the same radial distance than they do in ordinary boson stars.

\end{abstract}


\maketitle
\newpage

\section{INTRODUCTION}\label{sec: introduction}
    Boson stars (BSs) are localized self-gravitating lumps of scalar or vector fields~\cite{Schunck:2003kk,Liebling:2012fv}. As exotic horizonless compact objects, they have been considered as dark matter candidates in our Universe~\cite{Suarez:2013iw,Eby:2015hsq,Chen:2020cef} and potential mimickers of black holes (BHs)~\cite{Guzman:2009zz,Bambi:2025wjx,Macedo:2026jfv}. In astrophysics, gravitational-wave signals from boson star mergers can resemble those from binary black hole mergers~\cite{Palenzuela:2006wp,Cardoso:2016oxy,Palenzuela:2017kcg,Helfer:2018vtq,Croon:2018ftb,Sanchis-Gual:2018oui,Evstafyeva:2024qvp,Ge:2024itl,Evstafyeva:2026juq} and are used to interpret some real gravitational-wave events~\cite{CalderonBustillo:2020fyi,CalderonBustillo:2022cja,ET:2025xjr}. In addition, BSs can produce (effective) shadows~\cite{Olivares:2018abq,Herdeiro:2021lwl,Rosa:2022tfv} and gravitational-lensing signatures~\cite{Dabrowski:1998ac,Sengo:2022jif,Yang:2025yej} that may resemble, but can also differ from, those of black holes.

    Historically, the study of boson stars can be traced back to the mid-1950s, when J.~A.~Wheeler sought to construct stable, particle-like configurations from smooth classical electromagnetic fields coupled to general relativity. However, the resulting solutions, known as ``geons", were unstable~\cite{Wheeler:1955zz}. More than a decade later, D.~J.~Kaup replaced the electromagnetic field with a free complex scalar field and obtained Klein–Gordon geon solutions, which later became known as mini boson stars~\cite{Kaup:1968zz}. At about the same time, Ruffini and Bonazzola studied a system of self-gravitating bosons from a quantum field theory perspective, showing that such boson stars can be understood as macroscopic quantum states composed of bosons occupying the same ground state, namely Bose–Einstein condensates~\cite{Ruffini:1969qy}. Similar to the quantum states of an atom, boson stars can also be classified into a ground state and different excited states according to the number of nodes in the scalar field function. 

    Boson star research now extends beyond the original mini boson star model and encompasses a variety of theoretical models. In addition to finding different solutions within the mini boson star model, there are two commonly considered approaches to extending this model. The first is to consider modified theories of gravity, including Gauss-Bonnet gravity~\cite{Hartmann:2013tca}, Palatini $f(R)$ gravity~\cite{Maso-Ferrando:2023nju}, $f(T)$ gravity~\cite{Ilijic:2020vzu}, and higher-dimensional spacetimes~\cite{Blazquez-Salcedo:2019qrz}. The second is to modify the matter-field sector. Within this approach, introducing self-interactions of the scalar field is an early and direct extension of the mini boson star model. In the literature, the commonly considered self-interaction potentials include quartic~\cite{Colpi:1986ye,Ryan:1996nk,Kling:2017hjm}, sextic~\cite{Kleihaus:2005me,Herdeiro:2021mol}, solitonic~\cite{Friedberg:1986tq,Grandclement:2014msa}, and axion potentials~\cite{Guerra:2019srj,Delgado:2020udb}. It has been shown that suitable self-interactions can increase the maximum compactness~\cite{Amaro-Seoane:2010pks} of boson stars and affect their stability~\cite{Sanchis-Gual:2021phr,Ildefonso:2023qty,Brito:2025rld}.

    In addition to being coupled to a single scalar field, Einstein gravity can also be coupled to multiple matter fields. In 2003, S.~H.~Hawley and M.~W.~Choptuik considered two real scalar fields with a relative phase shift in time and constructed multi-field boson star solutions~\cite{Hawley:2002zn}. Subsequently, A.~Bernal et al. studied multistate boson stars composed of a ground state and a first excited state~\cite{Bernal:2009zy}. They found that these BSs have a higher critical mass than the ground-state BSs and are more stable than just the excited state. Another class of multi-field configurations is $\ell$-boson stars~\cite{Alcubierre:2018ahf,Sanchis-Gual:2021edp}. They consist of $2\ell+1$ complex scalar fields with the same radial profile but different magnetic quantum numbers. Although each field has an angular dependence, their total energy-momentum tensor remains spherically symmetric.

    Charged boson stars~\cite{Jetzer:1989av,Collodel:2019ohy}, formed by coupling a complex scalar field to the Maxwell field, are another noteworthy extension. In the Newtonian limit, the balance between gravitational attraction and Coulomb repulsion defines a critical charge $q_{\mathrm{Nc}}$. When the charge $q$ of the scalar particles exceeds $q_{\mathrm{Nc}}$, Coulomb repulsion overcomes gravitational attraction, preventing the particles from forming a gravitationally bound state. However, numerical studies of this model have found static, localized solutions even for $q>q_{\mathrm{Nc}}$, within a finite range of the charge parameter~\cite{Pugliese:2013gsa,Lopez:2023phk}.

    Recently, BSs coupled to nonlinear electrodynamics (NED) have also received attention. The original motivation for NED came from the effort to remove the divergences associated with point charges and their infinite self-energy. One well-known early example is Born-Infeld electrodynamics~\cite{Born:1934gh}. In this theory, the electric field of a point charge remains finite at the origin, while Maxwell electrodynamics is recovered in the weak-field limit. In 2023, V.~Jaramillo et al.~\cite{Jaramillo:2023lgk} studied the Einstein–Born–Infeld–scalar field system and obtained charged boson star solutions (see also in Ref.~\cite{DeBenedictis:2025hau}). They found that these solutions exhibit qualitative behavior similar to that of charged BSs in Maxwell electrodynamics, while differing quantitatively in properties such as their total mass and electric charge.
    
    Shortly afterward, the Bardeen~\cite{Wang:2023tdz} and Hayward NED models~\cite{Yue:2023sep,Chicaiza-Medina:2025wul} were also used to study BSs. These two NED models do not reduce to Maxwell electrodynamics in the weak-field limit, and their corresponding energy–momentum tensors can serve as matter sources for magnetically charged regular BHs (see Refs.~\cite{Lan:2023cvz,Carballo-Rubio:2025fnc} for recent overviews of regular BHs). From another perspective, these two boson star models can also be viewed as studies of the introduction of additional matter fields into magnetically charged regular spacetimes. Unlike the Born-Infeld boson stars, the resulting BS solutions with magnetic monopole charges employ a neutral complex scalar field. Moreover, in both the Bardeen and Hayward models, when the magnetic charge exceeds a critical value, a special class of solutions emerges with frequencies of the scalar field close to zero. For these special solutions, there is a special location $r_{\mathrm{ch}}$, near which the metric function $1/g_{rr}$ becomes very small, while $-g_{tt}$ is close to zero inside this location. Since for an observer at infinity, these properties of the metric imply that objects near $r_{\mathrm{ch}}$ move exceedingly slowly, as if ``frozen", these solutions are referred to as frozen boson stars (similar phenomena have also been discovered in various other models, e.g.,~\cite{Lue:1999zp,Lemos:2003gx,Stotyn:2013spa,Blazquez-Salcedo:2020crd,Brihaye:2025dlq,Chivers:2026fae}).
    
    These studies naturally raise the question of whether other NED models that support electrically charged regular BHs and reduce to Maxwell electrodynamics in the weak-field limit can, when coupled to a neutral complex scalar field, also give rise to BSs with distinctive properties. In other words, what would happen if scalar matter were further introduced into electrically charged regular spacetimes supported by such NED models? In this regard, the first exact electrically charged regular BH solution was sourced by the Ayón-Beato-García (ABG) NED proposed in 1998~\cite{Ayon-Beato:1998hmi}. In this paper, we consider the self-gravitating system consisting of a neutral complex scalar field and ABG NED, and obtain Ayón-Beato-García boson star (ABGBS) solutions with electric charge. These solutions exhibit a richer variety of properties. They include not only frozen BSs at relatively large electric charges, but also, at smaller electric charges, solutions that exhibit gravitational blueshift and possess a negative mass function in some regions. Additionally, the study of timelike orbits shows that trajectories in both types of special solutions differ significantly from those in ordinary boson stars.

    The paper is organized as follows. In Sec.~\ref{sec: model}, we provide the setup for ABGBSs. Numerical results for these BSs and the analysis of their physical properties are presented in Sec.~\ref{sec: results}. The conclusion and discussion are given in Sec.~\ref{sec: conclusion}.

\section{MODEL}\label{sec: model}
In this section, we first give a brief introduction to the model of Einstein-Klein-Gordon theory coupled to the nonlinear electrodynamics, whose action, in units where $c =\hbar=4\pi G= 1$, is as follows:
\begin{equation}
     S=\int \mathrm{d}^4 x \sqrt{-g}\left [\frac{R}{4}-\mathcal{L}^{(\rm EM)}-\mathcal{L}^{(\rm S)}\right],
         \label{eq:action}
\end{equation}
where $R$ is the Ricci curvature, and the Lagrangian of the scalar field is:
\begin{equation}
     \mathcal{L}^{(S)}=\left(\nabla^{\mu}\Phi \right)^*\left(\nabla_{\mu}\Phi \right)+\mu^2\Phi\Phi^*.
\end{equation}
$\mathcal{L}^{(EM)}$ represents the Lagrangian of the electromagnetic field and is a function of $\mathcal{F}=\frac{1}{4}F_{\mu\nu}F^{\mu\nu}$ with the electromagnetic field strength $F_{\mu\nu}=\partial_\mu A_{\nu}-\partial_\nu A_{\mu}$, in which $A_{\mu}$ is the electromagnetic potential. In general, it is very difficult~\cite{Fan:2016hvf} to explicitly express the Lagrangian density of the nonlinear electromagnetic field that constitutes the charged regular spacetime as a function of the electromagnetic field strength. Alternatively, the Lagrangian density of the electromagnetic field is usually described by another function obtained through the Legendre transformation (the so-called $P$ framework~\cite{Bronnikov:2000vy}) as follows~\cite{Salazar:1987ap}:
 \begin{equation}
\mathcal{H}^{\rm (EM)}=2\mathcal{F}\frac{\partial\mathcal{L}^{\rm (EM)}}{\partial \mathcal{F}}-\mathcal{L}^{\rm (EM)},
\end{equation}
Defining $P_{\mu\nu} = \frac{\partial\mathcal{L}^{\rm (EM)}}{\partial \mathcal{F}}F_{\mu\nu}$ and $P = \frac{1}{4}P_{\mu\nu}P^{\mu\nu}$,
one can find that $\mathcal{H}^{\rm (EM)}$ is a function of $P$ and $F_{\mu\nu}=\frac{\partial\mathcal{H}^{\rm (EM)}}{\partial P}P_{\mu\nu}$. For the Ayón-Beato-García NED proposed in 1998, $\mathcal{H}^{\rm (EM)}$ reads~\cite{Ayon-Beato:1998hmi}\footnote{Following Refs.~\cite{Wang:2023tdz,Yue:2023sep}, we redefine the parameter $s$: our parameter ``$s$" is equal to ``$sq^2$" in Ref.~\cite{Ayon-Beato:1998hmi}. This redefinition does not lead to any essential changes~\cite{Tan:2025jcg}.}
\begin{equation}
\mathcal{H}^{\rm (EM)}=P\frac{1-3\sqrt{-2q^2P}}{\left(1+\sqrt{-2q^2P}\right)^3}-\frac{3}{2s}\left(\frac{\sqrt{-2q^2P}}{1+\sqrt{-2q^2P}}\right)^{5/2},
\label{eq:H}
\end{equation}
where $q$ and $s$ are independent parameters. Then, the Lagrangian density $\mathcal{L}^{\rm (EM)}$ can be derived as
\begin{align}
\mathcal{L}^{\rm (EM)}&=2P\frac{\partial\mathcal{H}^{\rm (EM)}}{\partial P}-\mathcal{H}^{\rm (EM)}\\
&=P\frac{\left(1-8\sqrt{-2q^2P}-6q^2P\right)}{\left(1+\sqrt{-2q^2P}\right)^4}-\frac{3}{4s}\frac{\left(-2q^2P\right)^{5/4}\left(3-2\sqrt{-2q^2P}\right)}{\left(1+\sqrt{-2q^2P}\right)^{7/2}}.
\end{align}

It is worth noting that, for $P\to0$, Eq.~\eqref{eq:H} gives
$\mathcal H^{\mathrm EM}\approx P$ and $\mathcal{H}^{\mathrm EM}_P\equiv\partial \mathcal H^{\mathrm EM}/\partial P\to1$, so the ABG model can reduce to Maxwell electrodynamics in the weak-field limit. Moreover, since the action (\ref{eq:action}) is invariant under global $U (1)$ transformations, the system possesses a conserved Noether current:
\begin{equation}
    j^{\nu}=i g^{\mu\nu} \left[\Phi^*\nabla_{\mu}\Phi-\Phi(\nabla_{\mu}\Phi^*) \right].
\end{equation}
This implies the existence of a Noether charge:
\begin{equation}
    Q=\frac{1}{4\pi}\int_{\mathcal{S}} j^{\nu}n_{\nu}\mathrm{d}V, \label{eq:charge}
\end{equation}
here, $\mathcal{S}$ is a spacelike hypersurface (with the volume element $\mathrm{d}V$), and $n_{\nu}$ is the unit normal vector of $\mathcal{S}$. This conserved charge corresponds to the total number of bosons in the star~\cite{Jetzer:1989av}.

The Einstein, nonlinear electrodynamic~\cite{Ayon-Beato:1998hmi}, and Klein–Gordon equations arising from action \eqref{eq:action} are 
\begin{align}
    R_{\mu\nu}
    - \frac{1}{2}g_{\mu\nu}R
    &= 8\pi G\left(T^{\rm(S)}_{\mu\nu}+T^{\rm(EM)}_{\mu\nu}\right),
    \label{eq:einsteinB} \\
    \nabla_{\mu}P^{\mu\nu}
    &= 0,
    \label{eq:equationMaxwellFB} \\
    \nabla^{\mu}\nabla_{\mu}\Phi
    &= \mu^2\Phi.
    \label{eq:equationKGB}
\end{align}
with 
\begin{align}
    T^{\rm (EM)}_{\mu \nu}&=\frac{\partial\mathcal{H}^{\rm (EM)}}{\partial P}P_{\mu\lambda}P_{\nu}^{\lambda}-g_{\mu\nu}\left(2P\frac{\partial\mathcal{H}^{\rm (EM)}}{\partial P}-\mathcal{H}^{\rm (EM)}\right),\\
    T^{\rm (S)}_{\mu \nu}&=\nabla_{\mu} \Phi^* \nabla_{\nu} \Phi + \nabla_{\mu} \Phi \nabla_{\nu} \Phi^* \nonumber - g_{\mu \nu} \left( \nabla^{\alpha} \Phi^* \nabla_{\alpha} \Phi + \mu^2 \Phi^* \Phi \right).
\end{align}

Restricting to static, spherically-symmetric solutions of the field equations, we employ the following ansatz
\begin{equation}
    \mathrm{d} s^2=-n(r) \sigma^2(r) \mathrm{d} t^2+\frac{\mathrm{d} r^2}{n(r)}+r^2\left(\mathrm{d} \theta^2+\sin ^2 \theta \mathrm{d} \varphi^2\right),
    \label{eq:metric}
\end{equation}
here, $n(r)$ and $\sigma(r)$ are functions of the radial coordinate $r$ only. Note that the metric component $g_{rr}=1/n(r)$ defines the radial mass function $m(r)$~\cite{Misner:1964je}:
\begin{equation}
    1/g_{rr}= n(r)=1-\frac{2m(r)}{r}.
\end{equation}
In the limit $r\to\infty$, the ADM mass of the BSs is given by $M=m(\infty)$.

For the electrodynamic field $A_{\mu}$, scalar field $\phi$ and auxiliary field $P_{\mu\nu}$, we assume
\begin{equation}
     A_{\mu}\mathrm{d}x^{\mu}=V(r)\mathrm{d}t,\quad  \Phi=\phi(r) e^{-\mathrm{i}\omega t},\quad P_{\mu\nu}= 2\delta^t_{[\mu} \delta^r_{\nu]}\frac{D(r)}{r^2},
    \label{eq:ansatz}
\end{equation}
where $\omega$ is the real constant corresponding to the frequency of oscillation of the scalar field, and the functions $V(r)$, $D(r)$ and $\phi(r)$ are three real radial functions. 

Due to the relationship $F_{\mu\nu}=\frac{\partial\mathcal{H}^{(EM)}}{\partial P}P_{\mu\nu}$, the potential functions $V(r)$ can be obtained from the function $D(r)$. In addition, from Eqs. \eqref{eq:charge}, \eqref{eq:metric}, \eqref{eq:ansatz}, the particle number is:
\begin{equation}
    Q= 2\omega \int^{\infty}_0 \frac{r^2\phi^2}{n\sigma}  \mathrm{d}r.
\end{equation} 
  Substituting the ansatz into its equations of motion and into the relationship $F_{\mu\nu}=\frac{\partial\mathcal{H}^{\rm(EM)}}{\partial P}P_{\mu\nu}$ yields a set of reduced equations about $D(r)$, $V(r)$, $\sigma(r)$, $\phi(r)$, and $n(r)$:
\begin{align}
D'&=\frac{\sigma'}{\sigma}D,\label{eq:D}\\
V'&=-Dr^{4}\left[
\frac{r^{2}-5q^{2}}{(r^{2}+q^{2})^{4}}
+\frac{15q^{3}}{4s(r^{2}+q^{2})^{7/2}}
\right],\label{eq:V}\\
\sigma'&=2r\sigma\left(
\phi'^{\,2}+\frac{\omega^{2}\phi^{2}}{n^{2}\sigma^{2}}
\right),\\
\phi''&=
-\left(\frac{2}{r}+\frac{n'}{n}+\frac{\sigma'}{\sigma}\right)\phi'
+\left(\frac{\mu^{2}}{n}
-\frac{\omega^{2}}{n^{2}\sigma^{2}}\right)\phi,\\
n'&=\frac{1-n}{r}
+\frac{q^{2}r(3q^{2}-r^{2})}{(r^{2}+q^{2})^{3}}
-\frac{3q^{5}r}{s(r^{2}+q^{2})^{5/2}}
-2r\left[
n\phi'^{\,2}
+\left(\mu^{2}+\frac{\omega^{2}}{n\sigma^{2}}\right)\phi^{2}
\right],
\label{eq:n}
\end{align}
where ``$\prime$"  denotes differentiation with respect to $r$.

It is worth noting that the action \eqref{eq:action} has two kinds of special solutions. When $q=0$, this model describes mini BSs. In contrast, in the case of vanishing complex scalar field, the solution of this model describes ABG solutions with electric charges. The metric is 
\begin{equation}
\mathrm{d} s^2=-f(r) \mathrm{d} t^2+f(r)^{-1} \mathrm{d} r^2+r^2\left(\mathrm{d} \theta^2+\sin ^2\theta\,\mathrm{d} \varphi^2\right),
\label{eq:SABG}
\end{equation}
with 
\begin{equation}
    f(r)=1+\frac{C}{r}-\frac{q^3r^2}{s(r^2+q^2)^{3/2}}+\frac{q^2r^2}{(r^2+q^2)^2}.
\end{equation}
If $C\neq0$, the solution contains a singularity. We refer to this solution simply as the ``singular ABG solution". When the parameters $s$, $q$, and the constant $C$ are appropriate, this solution can correspond to a BH. Furthermore, in the case where the outermost root $r_{\rm{out}}$ of $f(r)=0$ satisfies $f'(r_{\rm{out}})=0$, this BH corresponds to an extremal BH. From the asymptotic behavior of the metric function $f(r)$ at infinity, one can deduce the mass of this solution as $M=q^3/(2s)-C/2$. On the other hand, if one requires $C=0$, the solution~(\ref{eq:SABG}) reduces to the regular ABG solution, with the metric function $f(r)$ taking the following form:
\begin{equation}
f(r)=1-\frac{q^3r^2}{s(r^2+q^2)^{3/2}}
+\frac{q^2r^2}{(r^2+q^2)^2}.
\end{equation} 
From equation $f(r)=0$, it can be found that there exists a value $\alpha_{\mathrm{s}}\approx0.31709$. When $q>\sqrt{s/\alpha_{\mathrm{s}}}$, the solution is a black hole solution with two distinct horizons, while for $q=\sqrt{s/\alpha_{\mathrm{s}}}$, the solution corresponds to an extremal black hole~\cite{Ayon-Beato:1998hmi}. In addition, the auxiliary function $D(r)$ has the simple form
$D(r)=q$.

In order to numerically solve the set of ordinary differential equations (ODEs) in Eqs.~\eqref{eq:D}-\eqref{eq:n}, we need to impose
appropriate boundary conditions on $n(r)$, $\sigma(r)$, $\phi(r)$, $V(r)$, and $D(r)$. Considering the assumption of regularity and asymptotic flatness, the metric functions $n(r)$ and $\sigma(r)$ satisfy:
\begin{equation}
    n(0)=1,\quad \sigma(0)=\sigma_0,\quad n(\infty)=1,\quad \sigma(\infty)=1.
    \label{eq:boundary}
\end{equation}
For the constant $\sigma_0$, its value is determined by solving the system of ordinary differential equations. Additionally, for the complex scalar field $\phi(r)$, the electric potential $V(r)$ and the function $D(r)$, we require the following boundary conditions
\begin{equation}
    \left.\frac{\mathrm{d} \phi}{\mathrm{d} r}\right|_{r=0}=0,\quad \phi(\infty)=0,\quad \left.\frac{\mathrm{d} V}{\mathrm{d} r}\right|_{r=0}=0,\quad V(\infty)=0,\quad D(\infty)=q.
\end{equation}
The boundary condition for $D(r)$ at infinity is chosen so that, when the scalar field vanishes, $D(r)=q$ and the solution can be reduced to the regular ABG solution. With these boundary conditions, Eqs.~\eqref{eq:D} and \eqref{eq:V} imply that, as
$r\to\infty$,
\begin{equation}
V'(r)=-\frac{q}{r^2}+o(r^{-2}),
\qquad
V(r)=\frac{q}{r}+o(r^{-1}).
\end{equation}
Thus, $q$ is the electric charge of the system.
\section{NUMERICAL RESULTS}\label{sec: results}
To facilitate numerical computation, we adopt the following scaling
transformations to render the variables dimensionless:
\begin{equation}
r\to\frac{r}{\mu},\quad
\omega\to\mu\omega,\quad
q\to\frac{q}{\mu},\quad
s\to\frac{s}{\mu^2},\quad
m\to\frac{m}{\mu},\quad
D\to\frac{D}{\mu}.
\end{equation}
This scaling transformation is equivalent to setting $\mu=1$ in numerical calculations. In addition, without loss of generality, we choose $s=0.2$ in the numerical implementation. As such, we have two input parameters $q$ and $\omega$. To enable numerical treatment over the entire spatial domain, we employ a compact radial coordinate $\bar{r} = r/(r + 1)$, which maps the semi-infinite region $[0, \infty)$ onto the unit interval $[0, 1]$. The solutions are found by using the finite element method. For a comprehensive discussion and comparison of numerical methods used to construct BS solutions, see Ref.~\cite{Gervalle:2026let}. After obtaining the numerical solution, the inverse transformation $r=\bar{r}/(1-\bar{r})$ can be used to replace the compactified $\bar{r}$ coordinate with the $r$ coordinate. For ordinary BSs, the number of grid points in the integration region $[0,1]$ is $1000$, while for frozen BSs and solutions with frequencies close to the frozen solution frequency, since the functions vary more dramatically, we appropriately increase the number of grid points according to computational requirements. The iterative method is the Newton-Raphson method, and the relative error for the numerical solutions in this paper is estimated to be below $10^{-5}$. Moreover, we perform the numerical calculations using our numerical code. See Refs.~\cite{Huang:2024rbg,Liu:2025jjf} for convergence analysis. 

	\begin{figure}[!htbp]
		\centering		
		\subfigure{  
			\includegraphics[width=7.1cm]{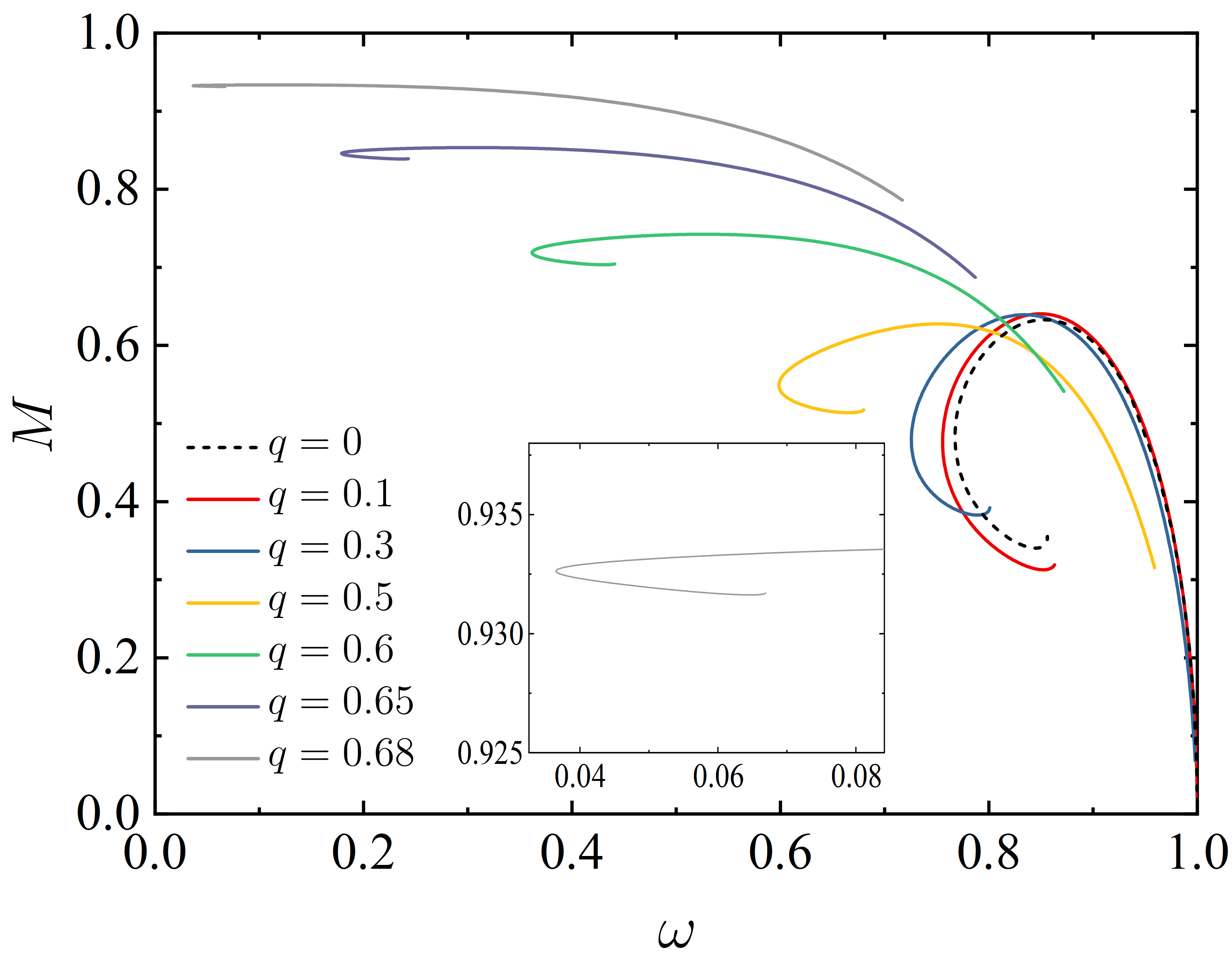}
		}
		\subfigure{
			\includegraphics[width=7.1cm]{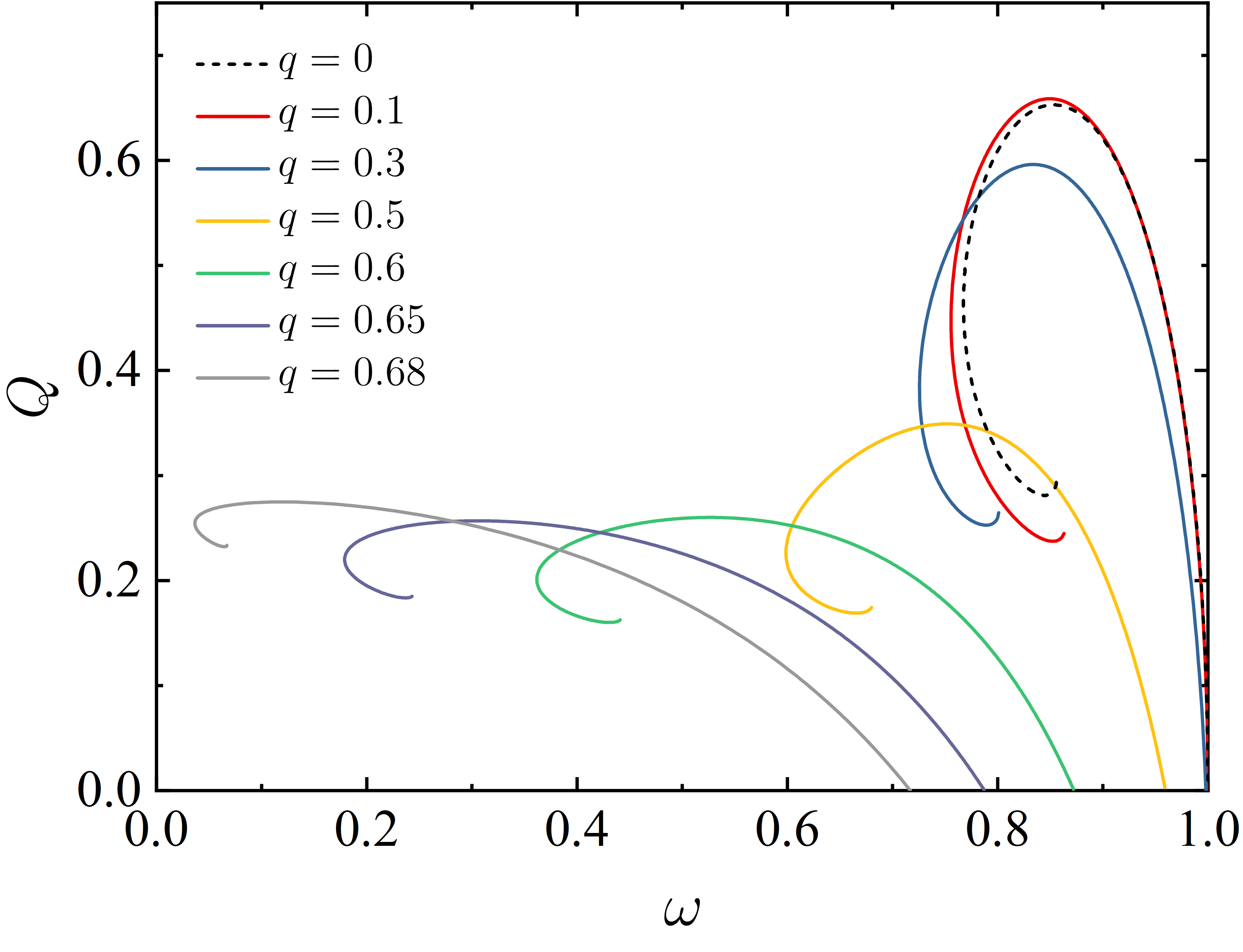}
   		} 	
        \subfigure{  
			\includegraphics[width=7.1cm]{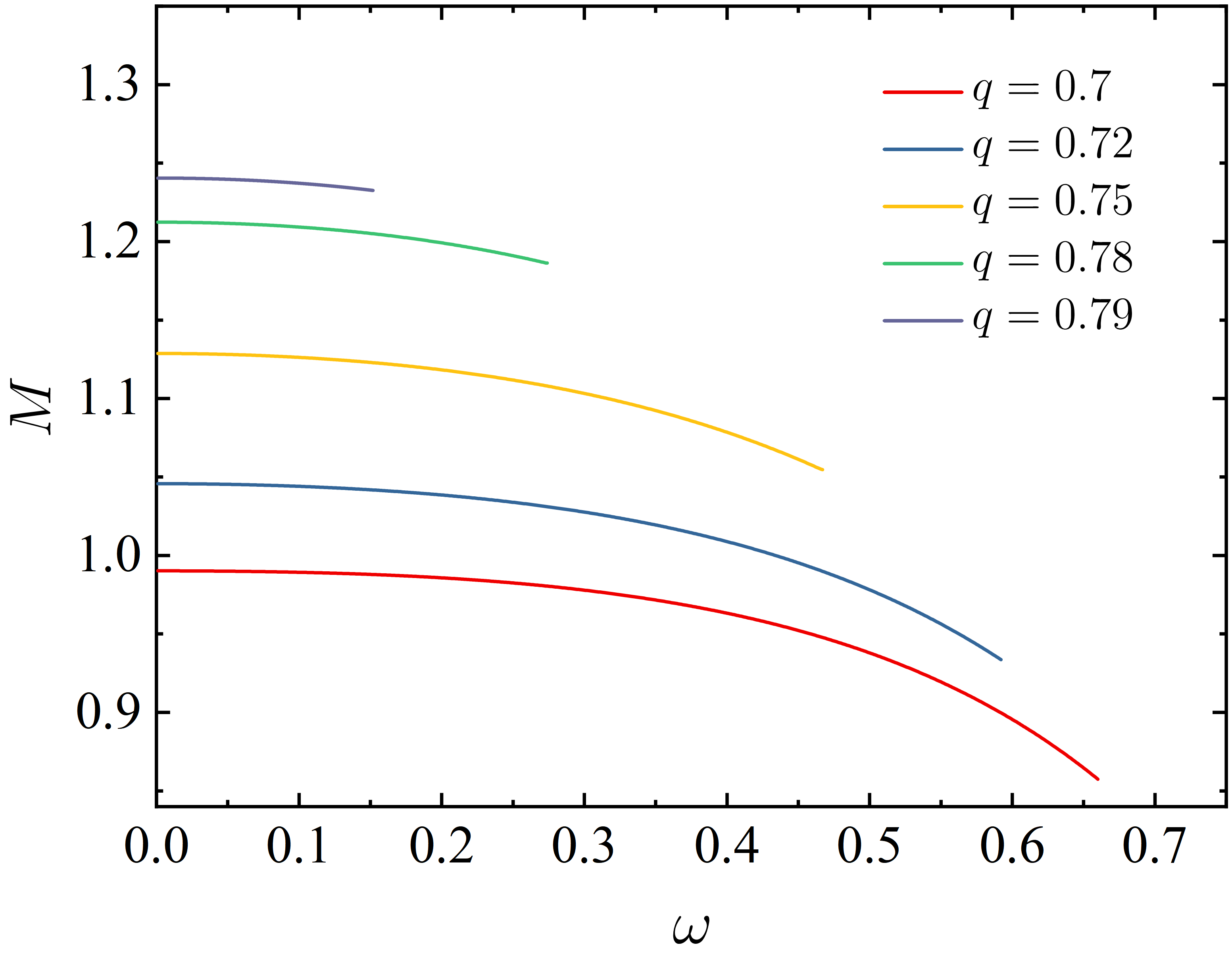}
		}
		\subfigure{
			\includegraphics[width=7.1cm]{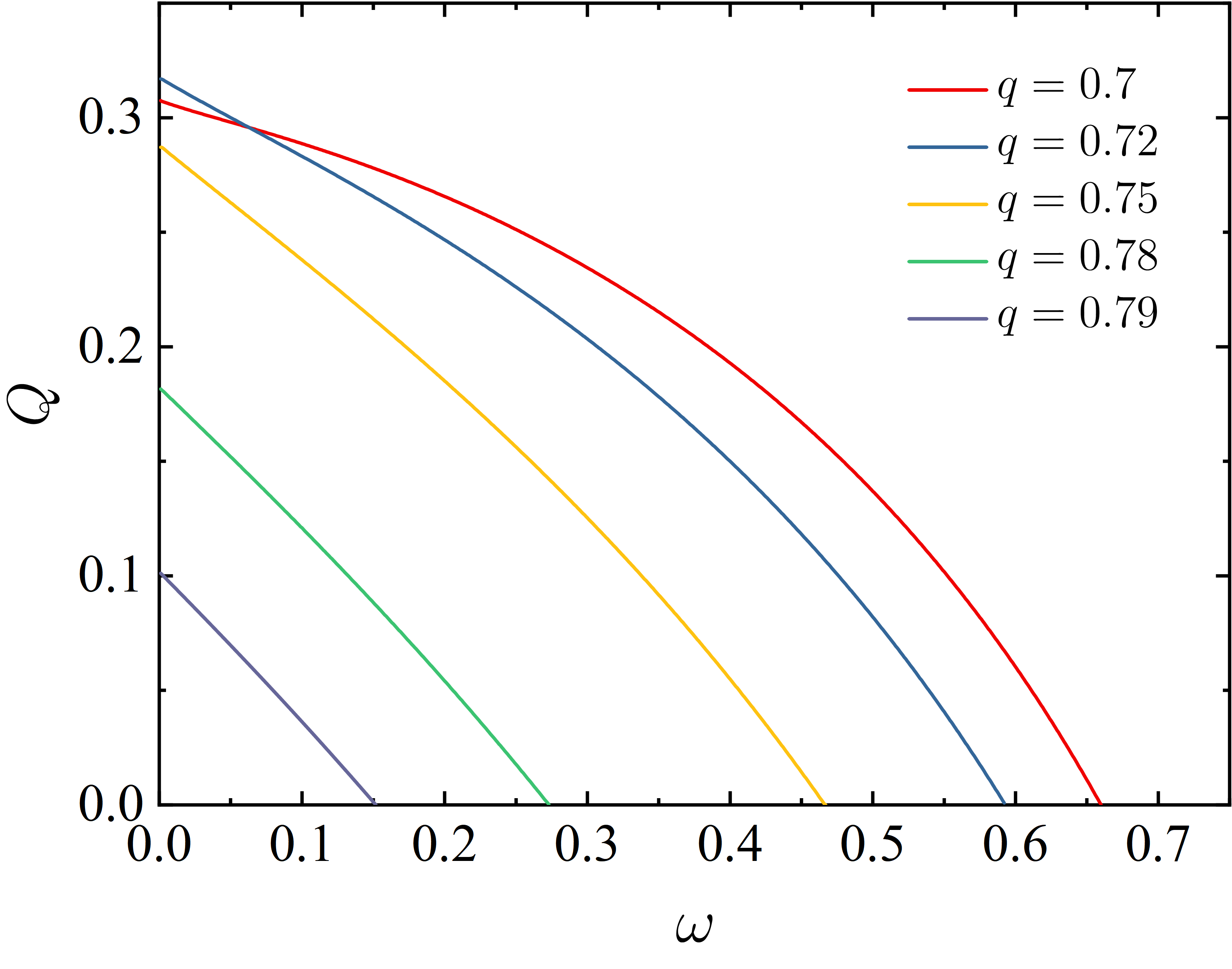}
   		} 	
		\caption{The ADM mass $M$ (left) and particle number $Q$ (right) as a function of frequency with different electric charges $q$. The top panels correspond to a smaller $q$, and the lower panels correspond to a larger $q$.}
    	    \label{fig:MQ}		
		\end{figure}

\begin{figure}[!htbp]
		\centering		
		\subfigure{  
			\includegraphics[height=.16\textheight, angle =0]{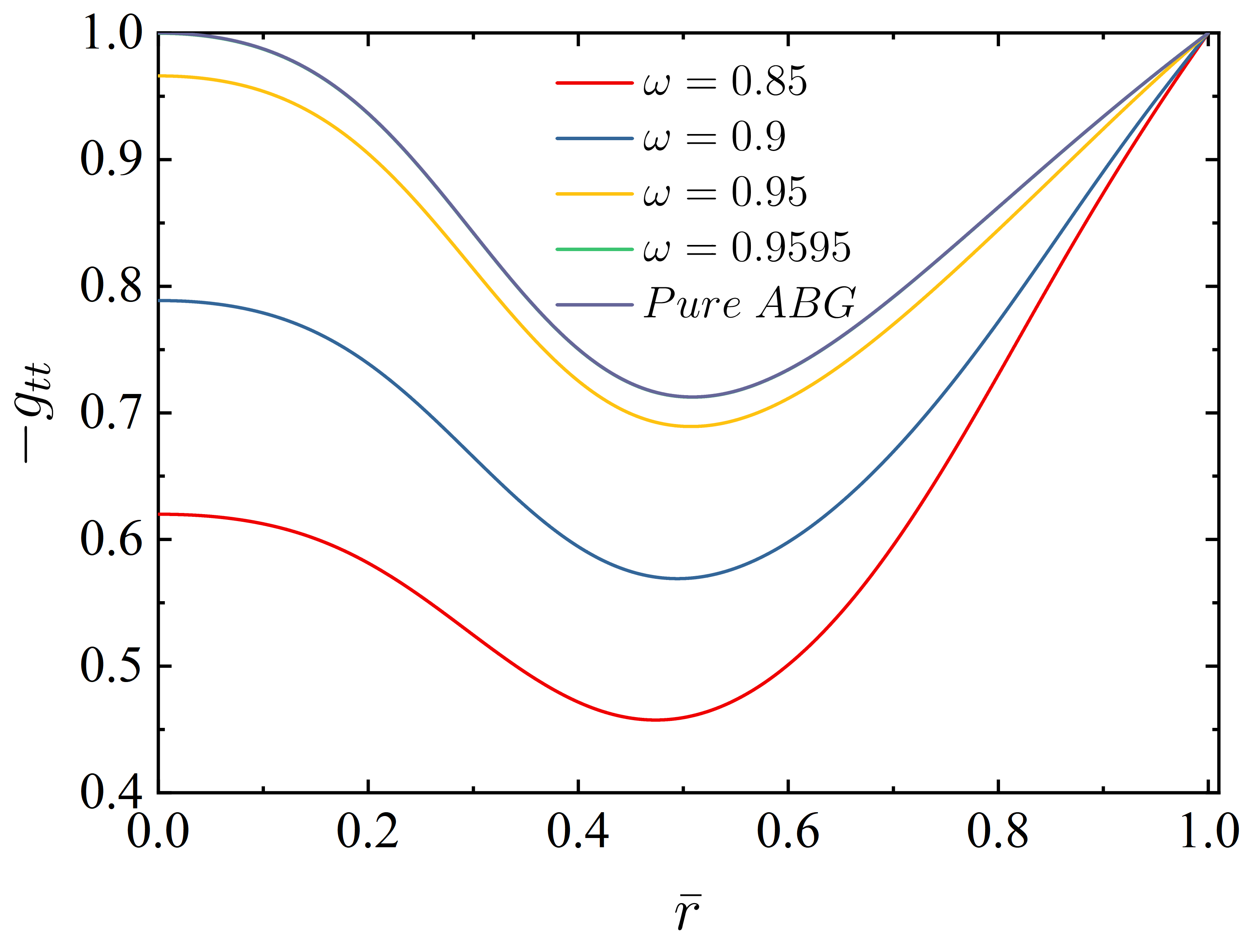}
		}
		\subfigure{
			\includegraphics[height=.16\textheight, angle =0]{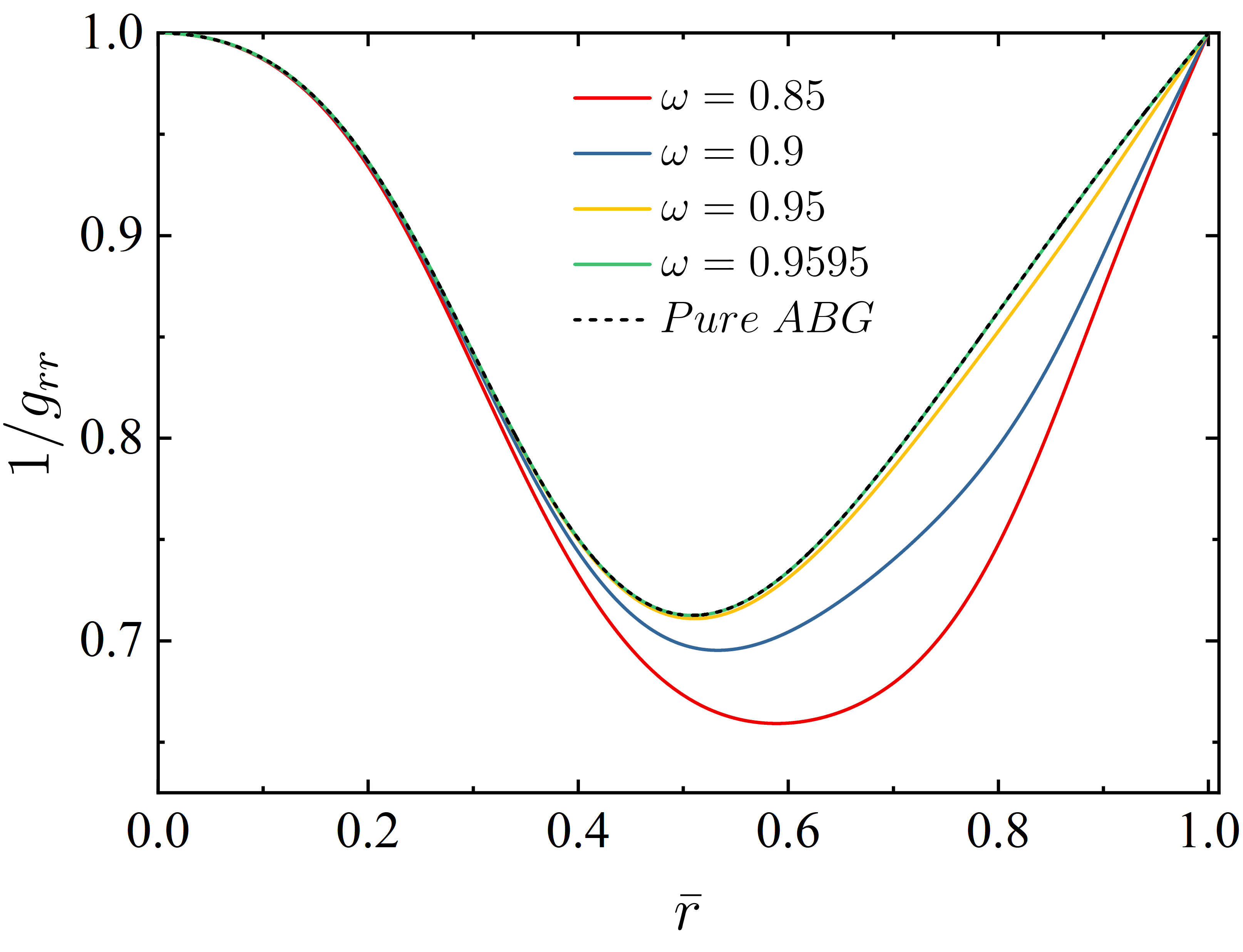}
   		} 	
     		\subfigure{
			\includegraphics[height=.16\textheight, angle =0]{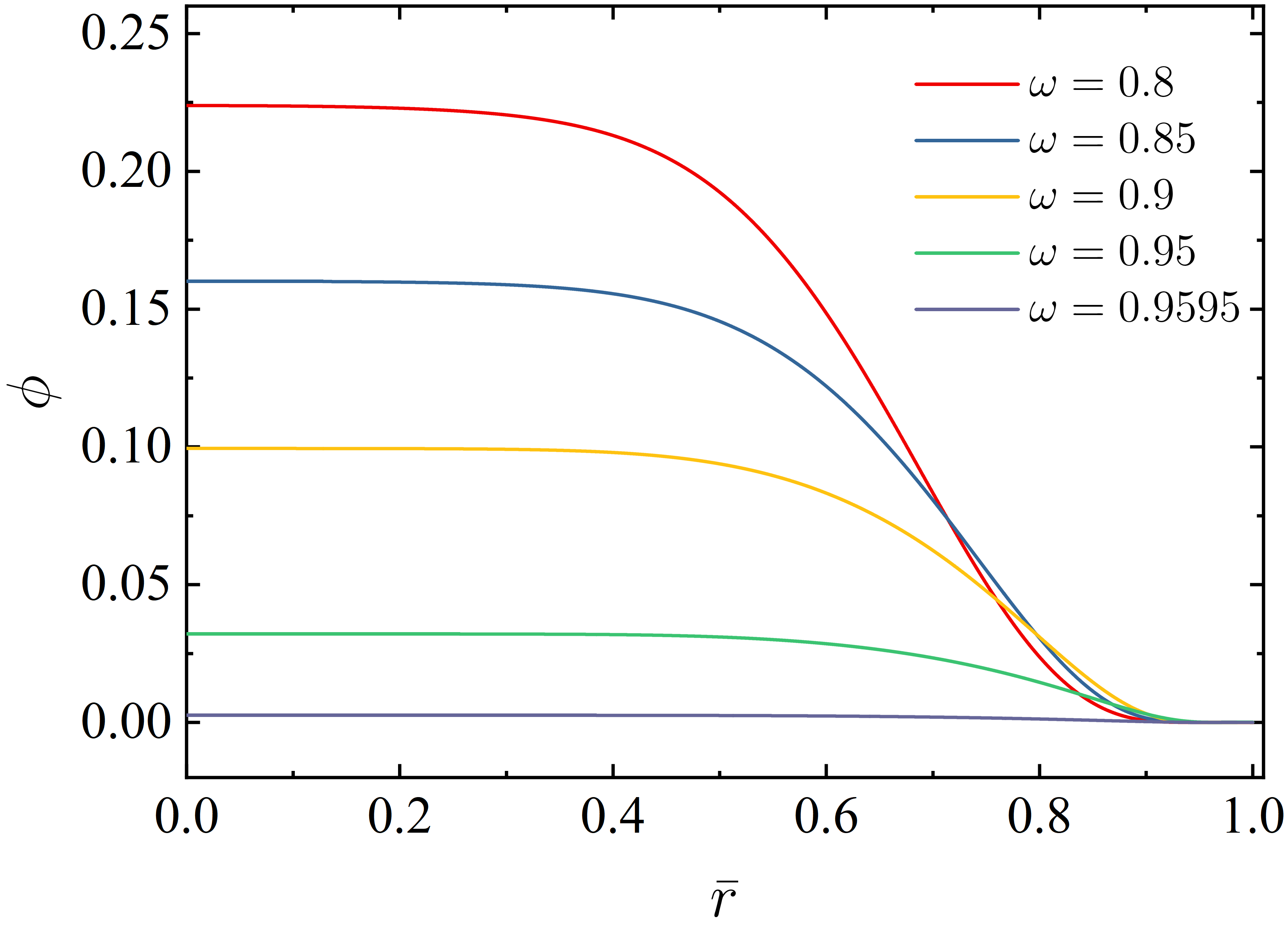}
   		} 	
		\caption{The radial profiles of the solutions for several relatively large values of the frequency, with $q=0.5$. For this value of $q$, the maximum frequency of the ABGBS solutions is approximately $0.9595$, at which the scalar field nearly vanishes.}
		\label{fig:vanish}		
		\end{figure}

Fig.~\ref{fig:MQ} exhibits the ADM mass $M$ (left) and particle number $Q$ (right) of ABGBSs versus
the frequency $\omega$ with the different values of electric charges $q$. These curves exhibit different features depending on the electric charge. For the case of $q=0$ (the mini BS), the $M(\omega)$ and $Q(\omega)$ curves start from $M = 0$ ($Q = 0$) for $\omega = 1$, in which limit the fields become very diluted and the solution trivialises. However, when $q\neq 0$, the maximum frequency of these curves is no longer equal to one, and the corresponding mass is also nonzero. In fact, as shown in Fig.~\ref{fig:vanish}, the solution in this case at the maximum frequency is the regular ABG solution.

Furthermore, for solutions with relatively small electric charges ($q<0.6863$, see top panels), the curves exhibit a spiral structure. Similar structures also arise in many other boson star models (see, e.g., Refs.~\cite{Alcubierre:2018ahf,Herdeiro:2017fhv,Herdeiro:2019mbz,Herdeiro:2020kvf,Herdeiro:2020jzx}). For any given one of these curves, the frequency $\omega$ of these solutions decreases from the maximum frequency to a minimum value, forming the first branch. Subsequently, the curve exhibits a backbending behavior, which gives rise to the second branch. Further calculations reveal additional branches. However, various stability analyses of many other boson star models have shown that stable solutions are generally confined to the first branch, whereas solutions on higher branches are typically unstable~\cite{Liebling:2012fv}. Therefore, unless otherwise stated, we compute the solutions only up to the second branch and present and analyze examples from the first branch.


\begin{figure}[!htbp]
		\centering		
		\subfigure{  
			\includegraphics[width=7.2cm]{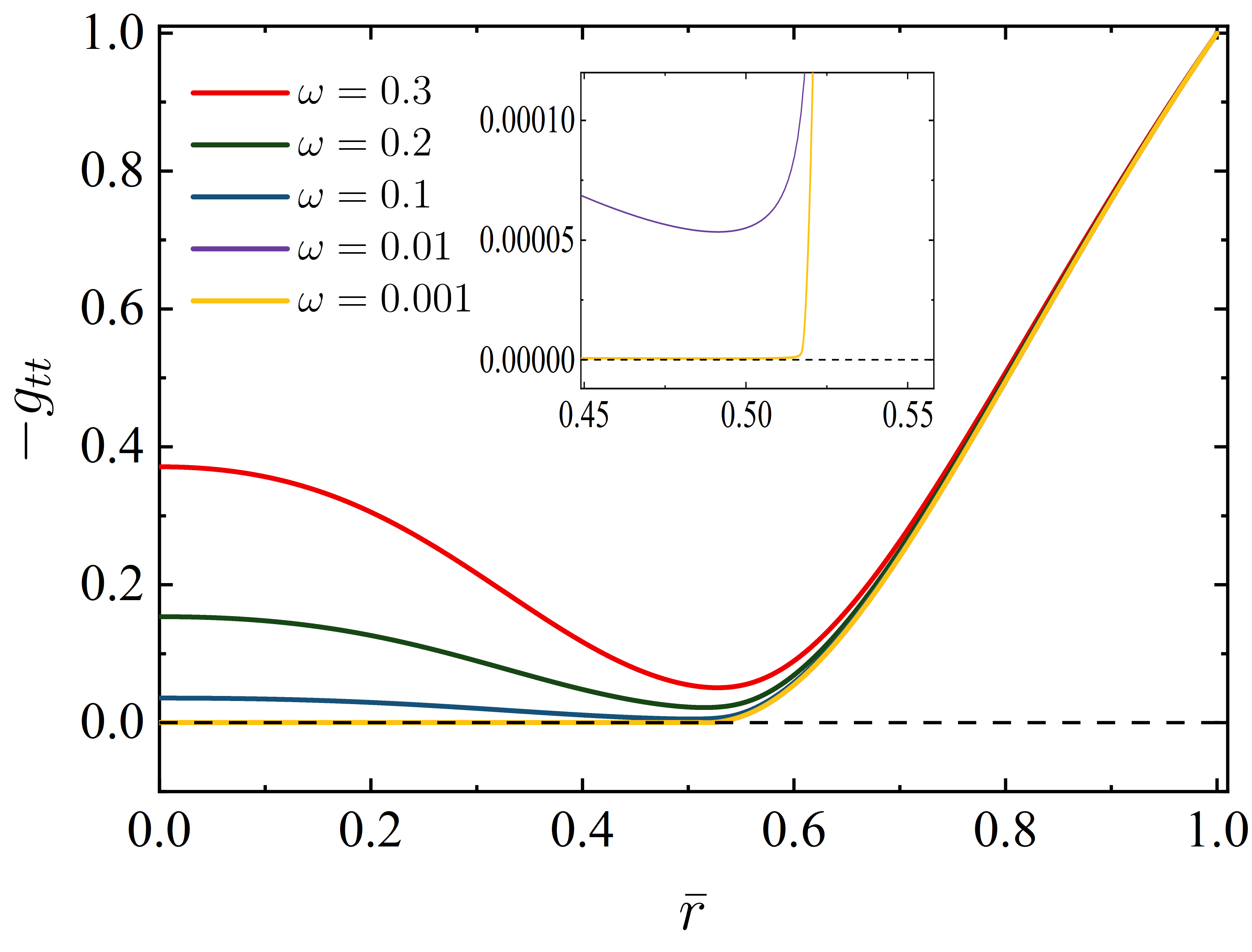}
		}
		\subfigure{
			\includegraphics[width=7.2cm]{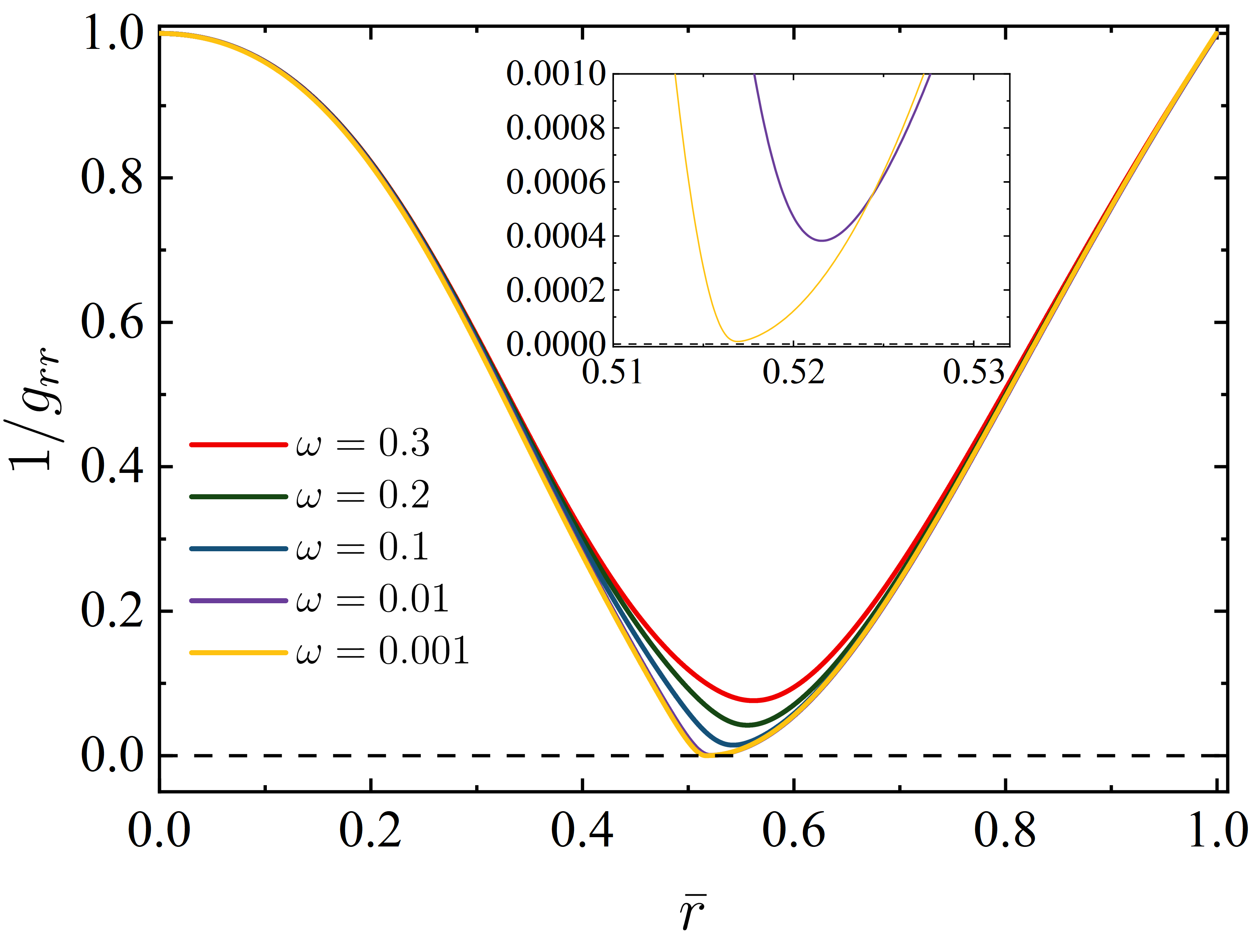}
   		} 
		\subfigure{  
			\includegraphics[width=7.2cm]{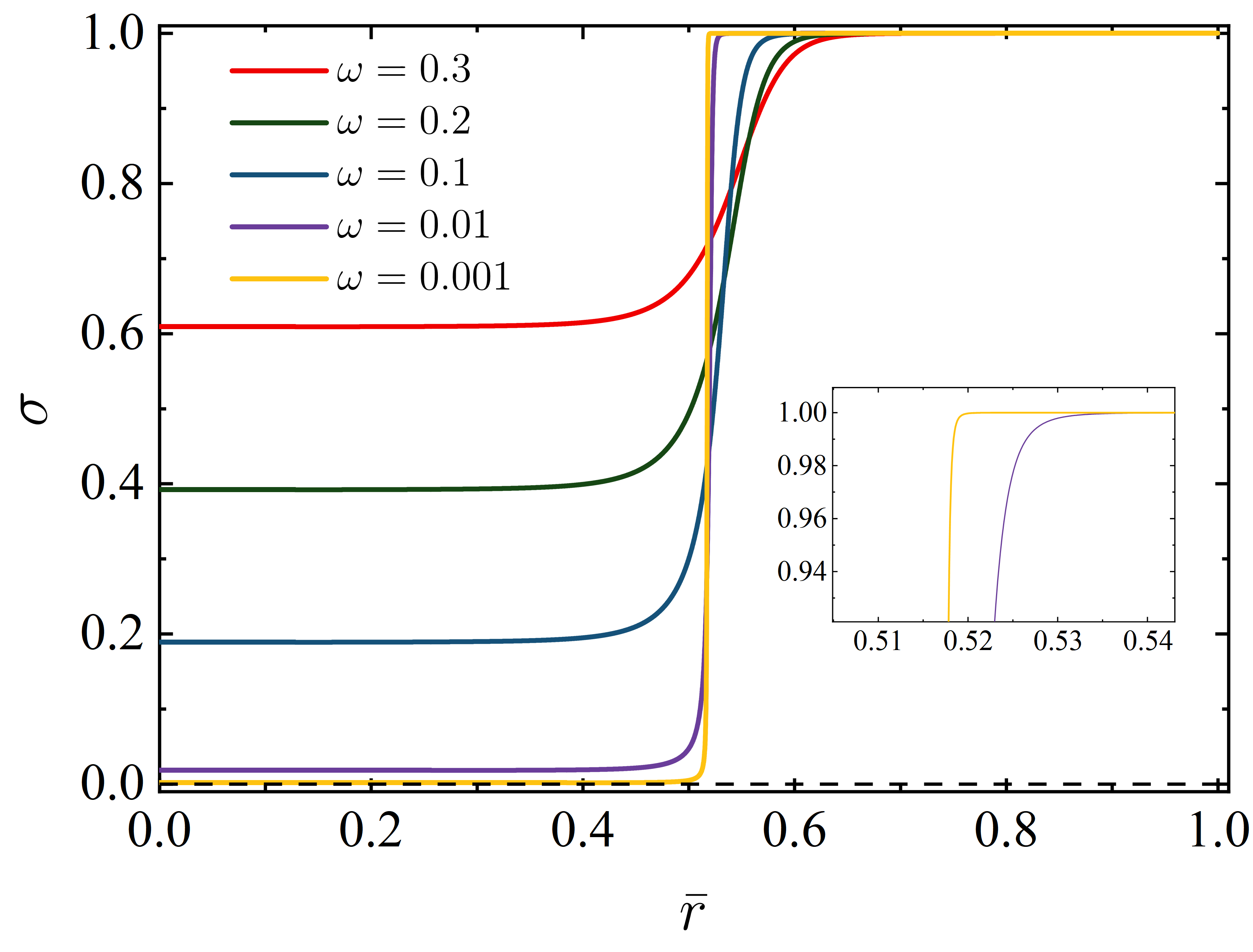}
		}
		\subfigure{
			\includegraphics[width=7.2cm]{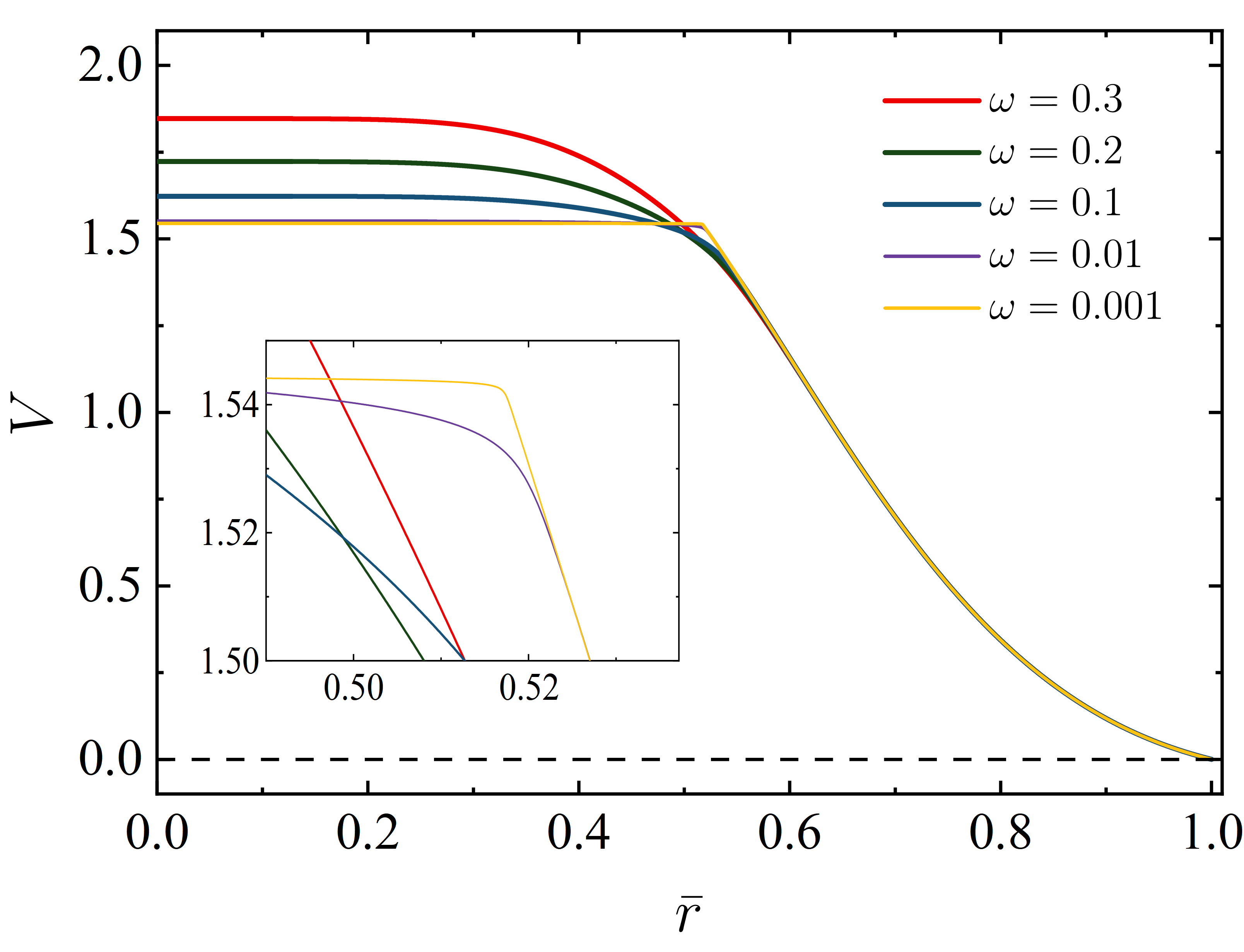}
   		} 
		\subfigure{  
			\includegraphics[width=7.2cm]{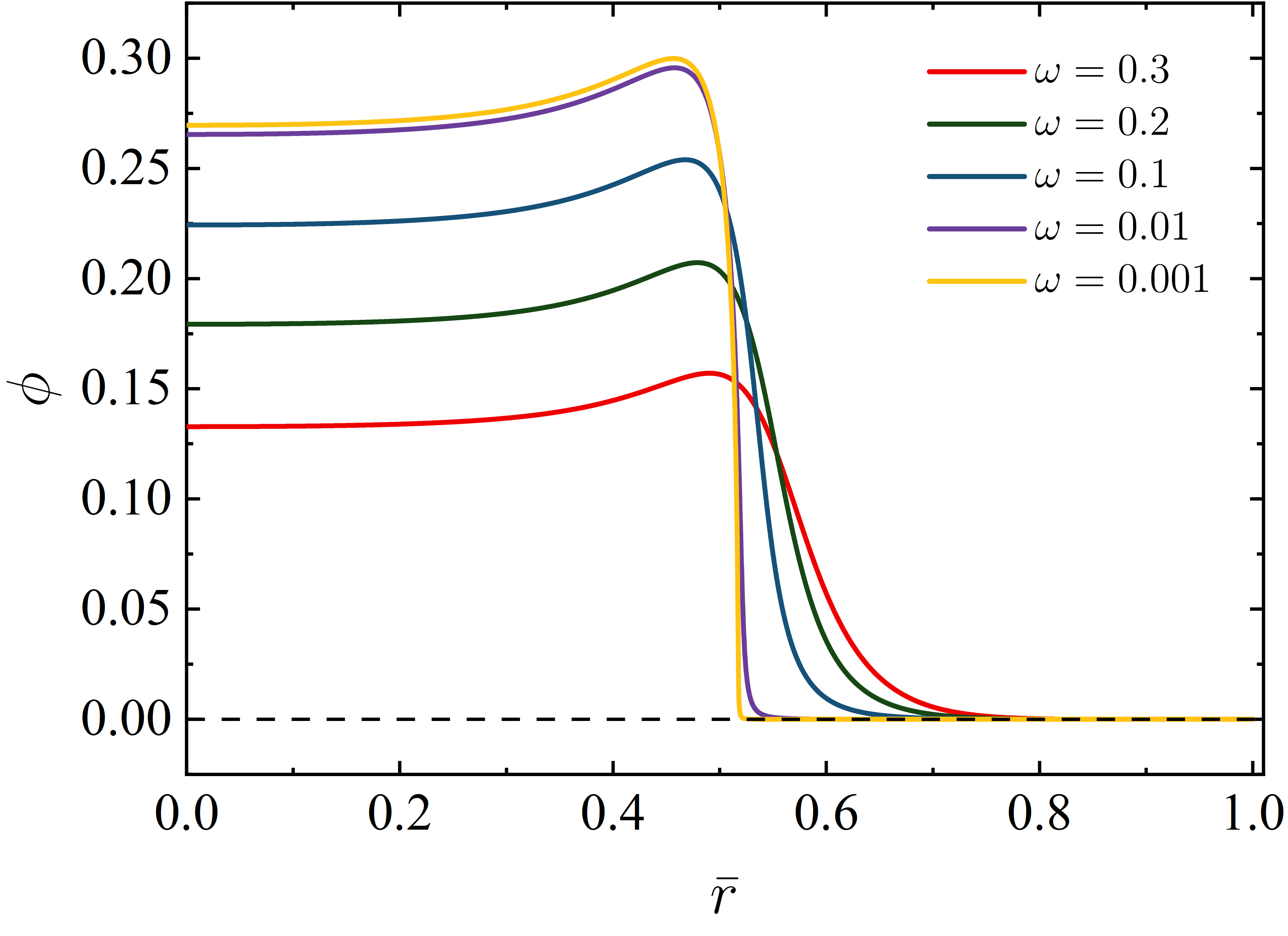}
		}
		\subfigure{
			\includegraphics[width=7.2cm]{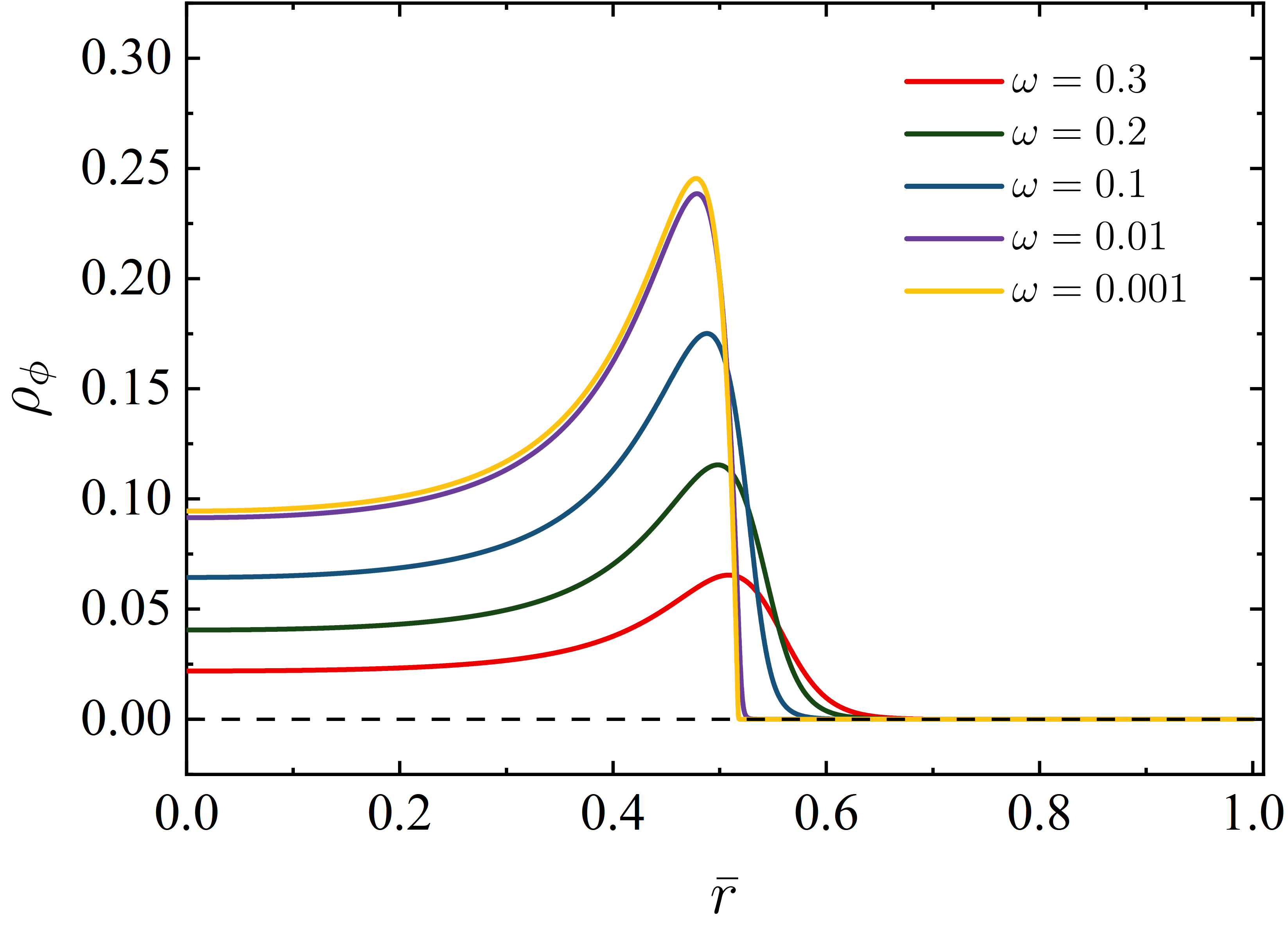}
   		}         
		\caption{The metric functions $-g_{tt}$, $1/g_{rr}$, and $\sigma(r)$, the electric potential $V$, the scalar field function $\phi$, and the energy density $\rho_{\phi}$ versus the compactified coordinate $\bar{r}$ for ABGBS solutions with several frequencies $\omega$ and an electric charge $q=0.75$.}
    	    \label{fig:BFfrozen}		
		\end{figure}

In contrast, for solutions with relatively large charges ($q>0.6863$, see bottom panels), the second branch vanishes, transforming the multi-valued curve into a single-valued curve, and the minimum frequency
can approach zero\footnote{We note that, as the grid resolution increases, the frequency $\omega$ can get closer and closer to zero. However, when $\omega=0$, we fail to find a solution within the allowable computational error. It is likely that no ABGBS solution exists for $\omega=0$. Moreover, given that the solution profiles for $\omega=0.01$ and $\omega=0.001$ nearly overlap (see Fig.~\ref{fig:BFfrozen}), and the solutions with a smaller frequency require substantial computational resources but exhibit no qualitative differences, we adopt the minimum value of the frequency as $0.001$.}.

\begin{figure}[!htbp]
		\centering		
		\subfigure{  
			\includegraphics[width=7.2cm]{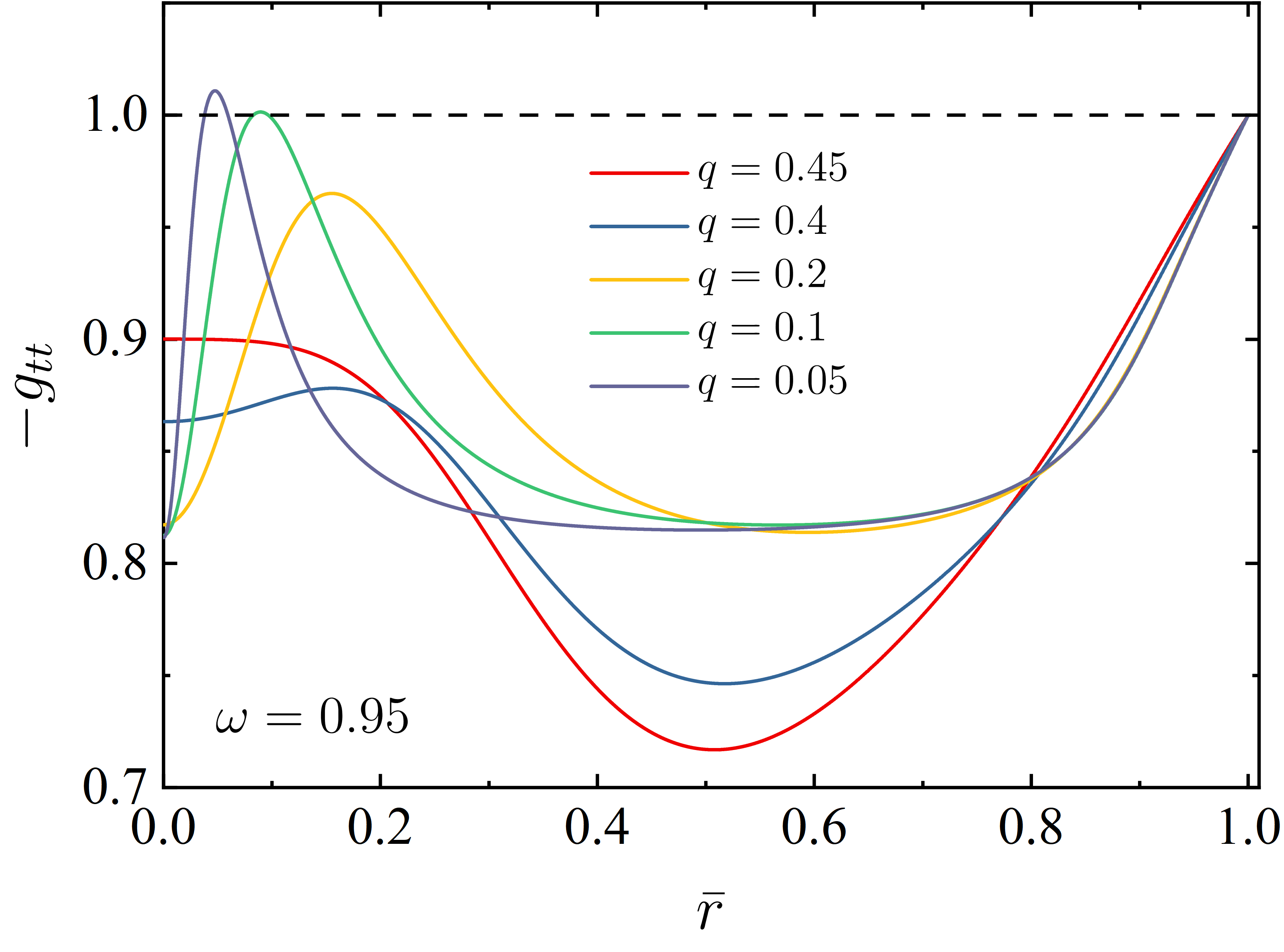}
		}
		\subfigure{
			\includegraphics[width=7.2cm]{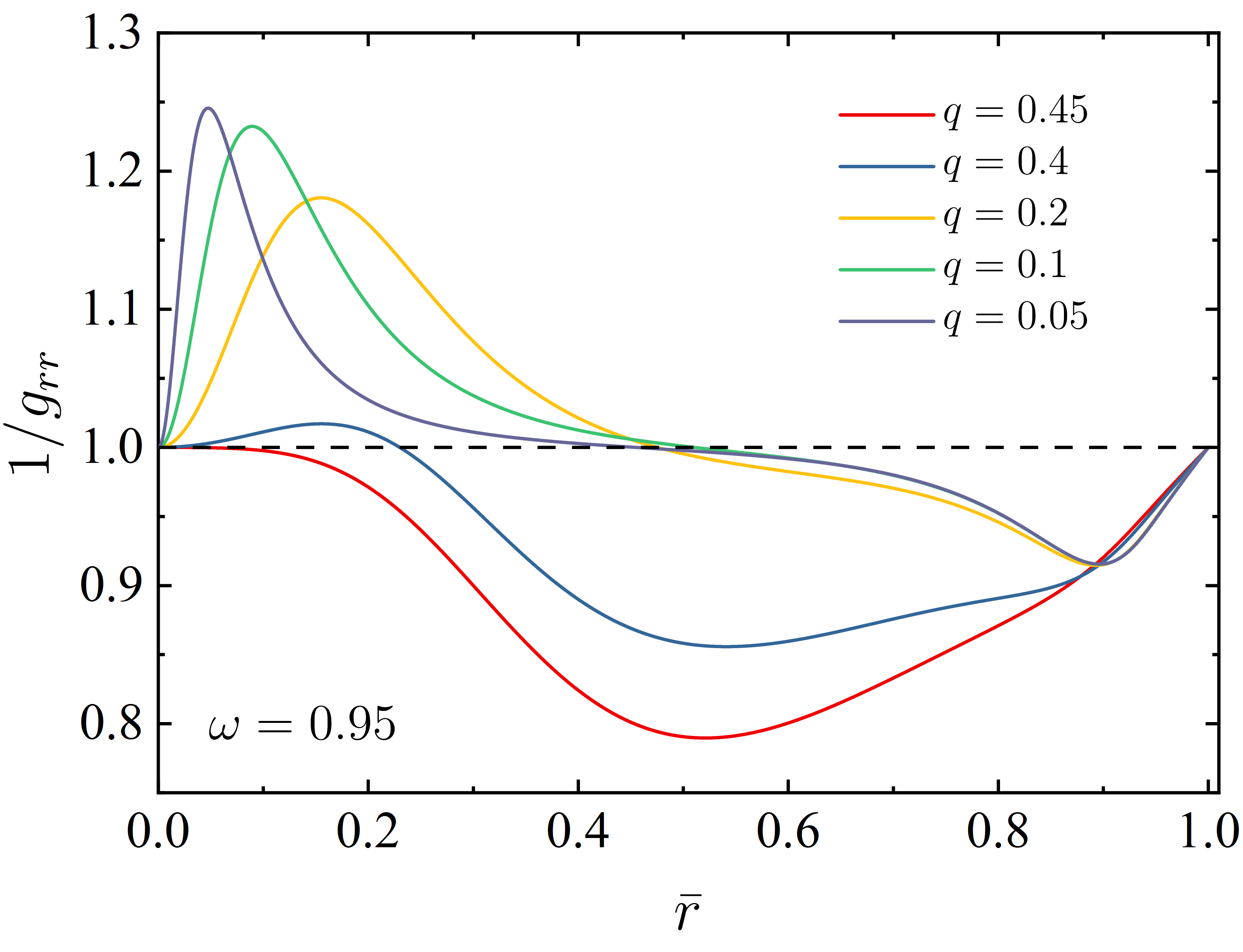}
   		} 	
        \subfigure{  
			\includegraphics[width=7.2cm]{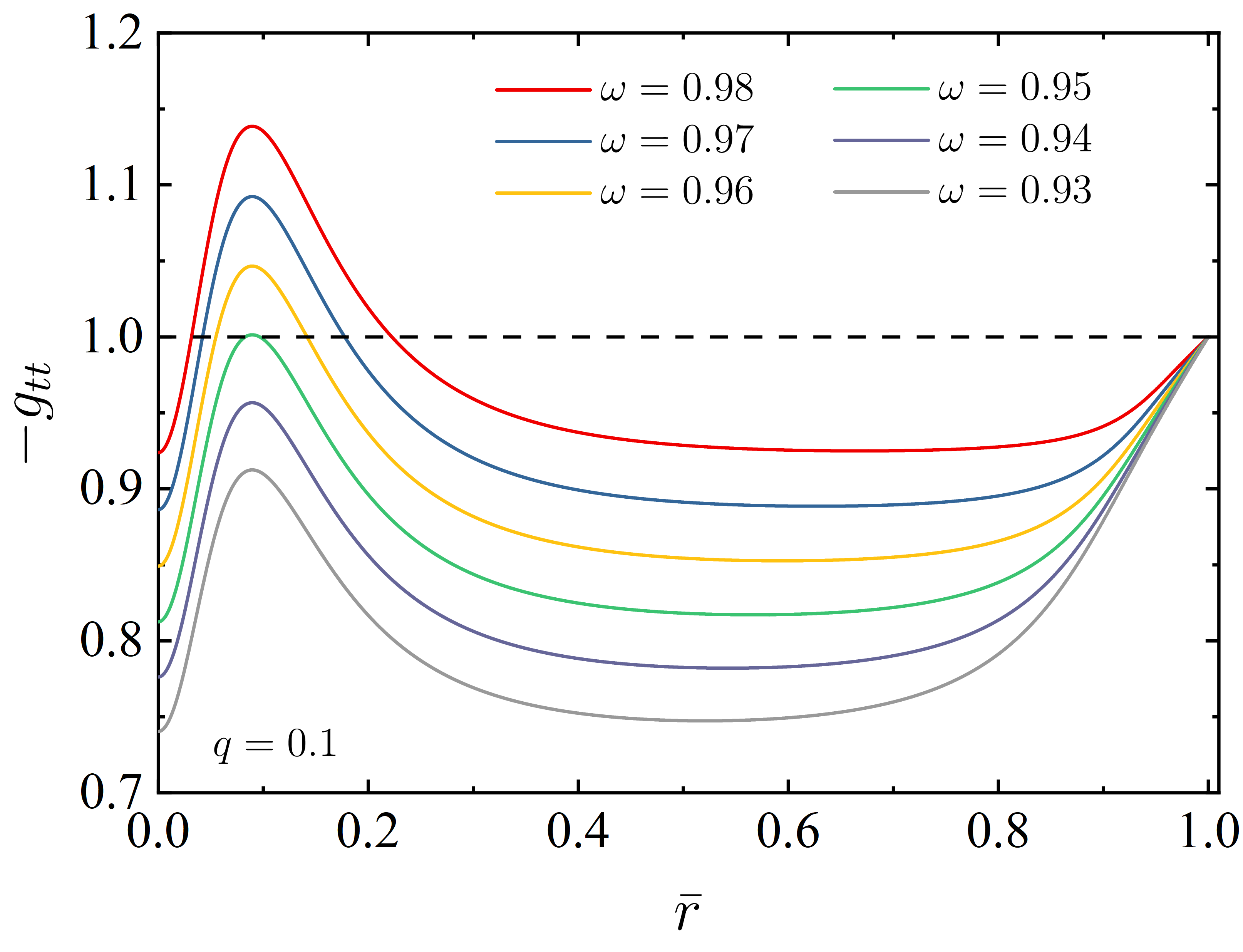}
		}
		\subfigure{
			\includegraphics[width=7.2cm]{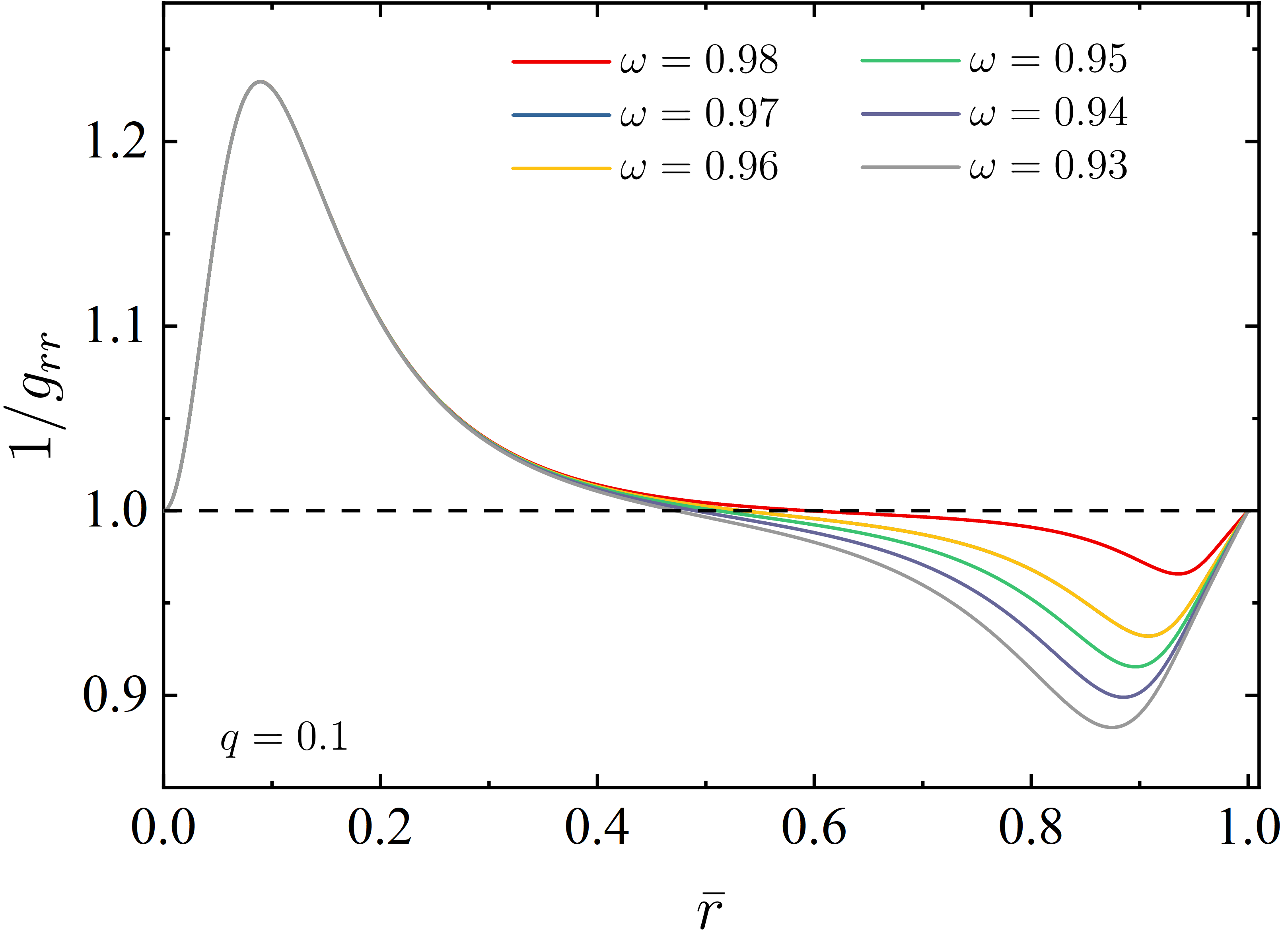}
   		} 	
		\caption{The radial profiles of the metric functions $-g_{tt}$ (left) and $1/g_{rr}$ (right) as functions of the compactified radial coordinate $\bar{r}=r/(1+r)$. The top panels show solutions with $\omega=0.95$ for different electric charges $q$, while the bottom panels show solutions with $q=0.1$ for different frequencies $\omega$. The horizontal dashed lines indicate unity.}
    	    \label{fig:BFfqk}		
		\end{figure}

The electric charge affects not only these global quantities of ABGBSs, but also the qualitative properties of the solutions. We first consider the case of small electric charges. As an example, Fig.~\ref{fig:BFfqk} shows the metric functions $-g_{tt}=n\sigma^2$ and $1/g_{rr}=n$ of the ABGBS solutions for a fixed charge of $q=0.1$ and a fixed frequency of $\omega=0.95$, respectively. For solutions with a fixed frequency $\omega=0.95$, both $-g_{tt}$ and $1/g_{rr}$ can exceed unity in certain regions when the electric charge is small. As the charge increases, $-g_{tt}$ no longer exceeds unity, whereas $1/g_{rr}$ can still exceed unity near the center. When the charge increases further, neither function exceeds unity. In other words, both metric functions exceed unity only when the frequency of the ABGBS solutions is close to its maximum value, corresponding to solutions close to the pure regular ABG solution (see Fig.~\ref{fig:vanish}).

In fact, for sufficiently small electric charges, the metric functions of the pure regular ABG solution, $-g_{tt}=1/g_{rr}=f(r)$, can exceed unity near the center. Expanding $f(r)$ around $r=0$, we obtain
\begin{equation}
f(r)=1+\left(\frac{1}{q^2}-\frac{1}{s}\right)r^2+O(r^4).
\label{eq:RABGF0}
\end{equation}
Therefore, $f(r)>1$ near the center when $q<\sqrt{s}$. In particular, for fixed $s$, the coefficient of $r^2$ increases as the charge $q$ decreases, so that $f(r)$ departs more rapidly from unity near the center.

	\begin{figure}[!htbp]
		\centering		
		\subfigure{
			\includegraphics[width=7.2cm]{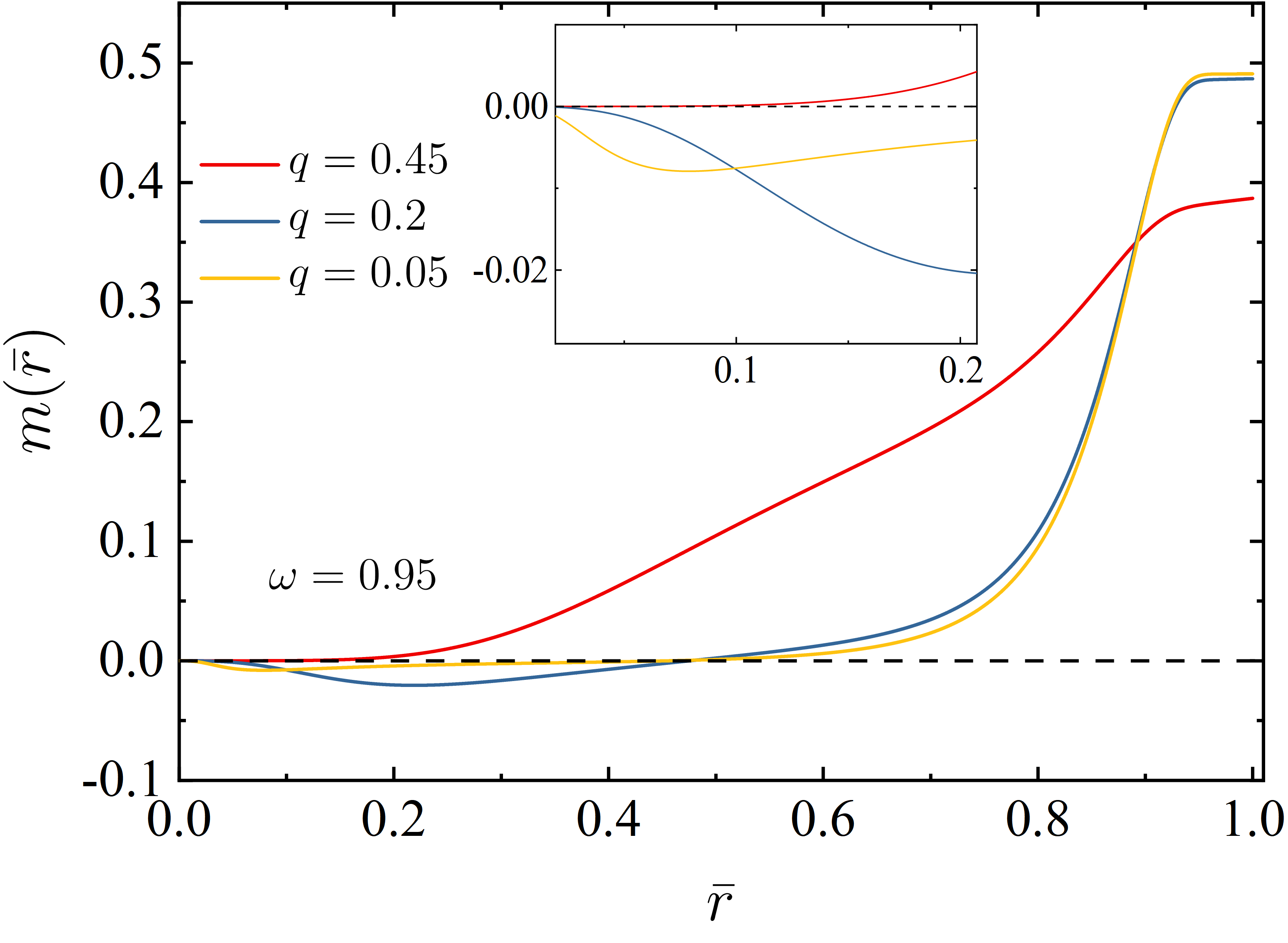}
   		} 		
		\caption{The mass function $m(\bar{r})$ versus the compactified coordinate $\bar{r}$ for different values of the charge $q$.}
    	    \label{fig:BFf}		
		\end{figure}

The metric functions $-g_{tt}$ and $1/g_{rr}$ of the ABGBS solutions can exceed unity in certain regions, corresponding to gravitational blueshift and a negative mass function, respectively. Fig.~\ref{fig:BFf} shows the mass function of the ABGBS solutions for different electric charges at $\omega=0.95$. For solutions with $1/g_{rr}>1$, such as those with $q=0.2$ and $q=0.05$, the mass function is negative near the center. However, it should be noted that the mass function approaches a positive value at infinity, implying that the ADM mass remains positive. The emergence of a negative mass function implies that the total energy density $\rho=-\left(T^{(\mathrm {S})t}{}_{t}+T^{(\mathrm{EM})t}{}_{t}\right)$ is negative in some regions, where the weak energy condition and the dominant energy condition are violated. Requiring these energy conditions to hold would consequently impose restrictions on the parameter space of ABGBS solutions.


	\begin{figure}[!htbp]
		\centering		
		\subfigure{  
			\includegraphics[width=7.2cm]{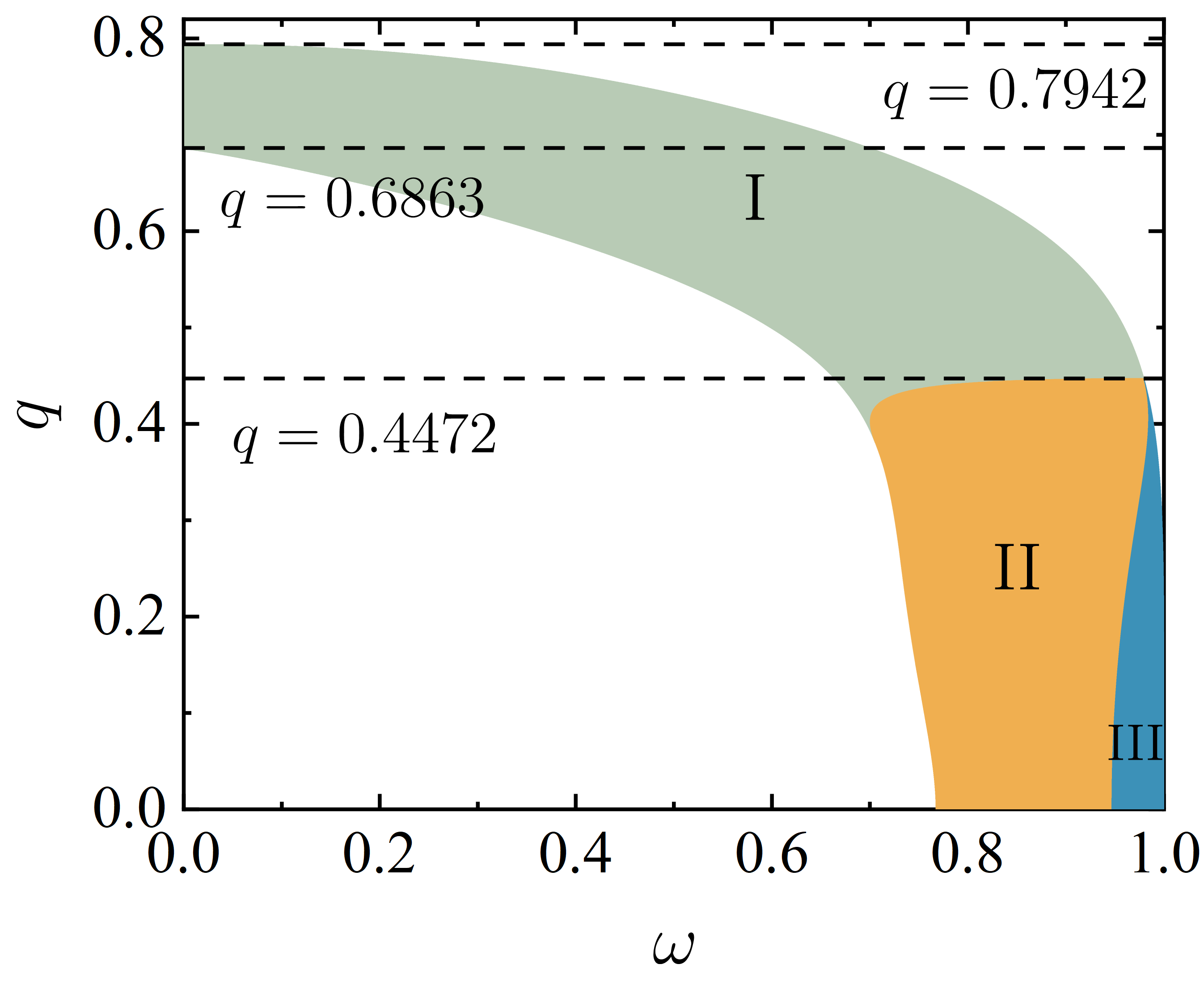}
		}
		\subfigure{
			\includegraphics[width=7.2cm]{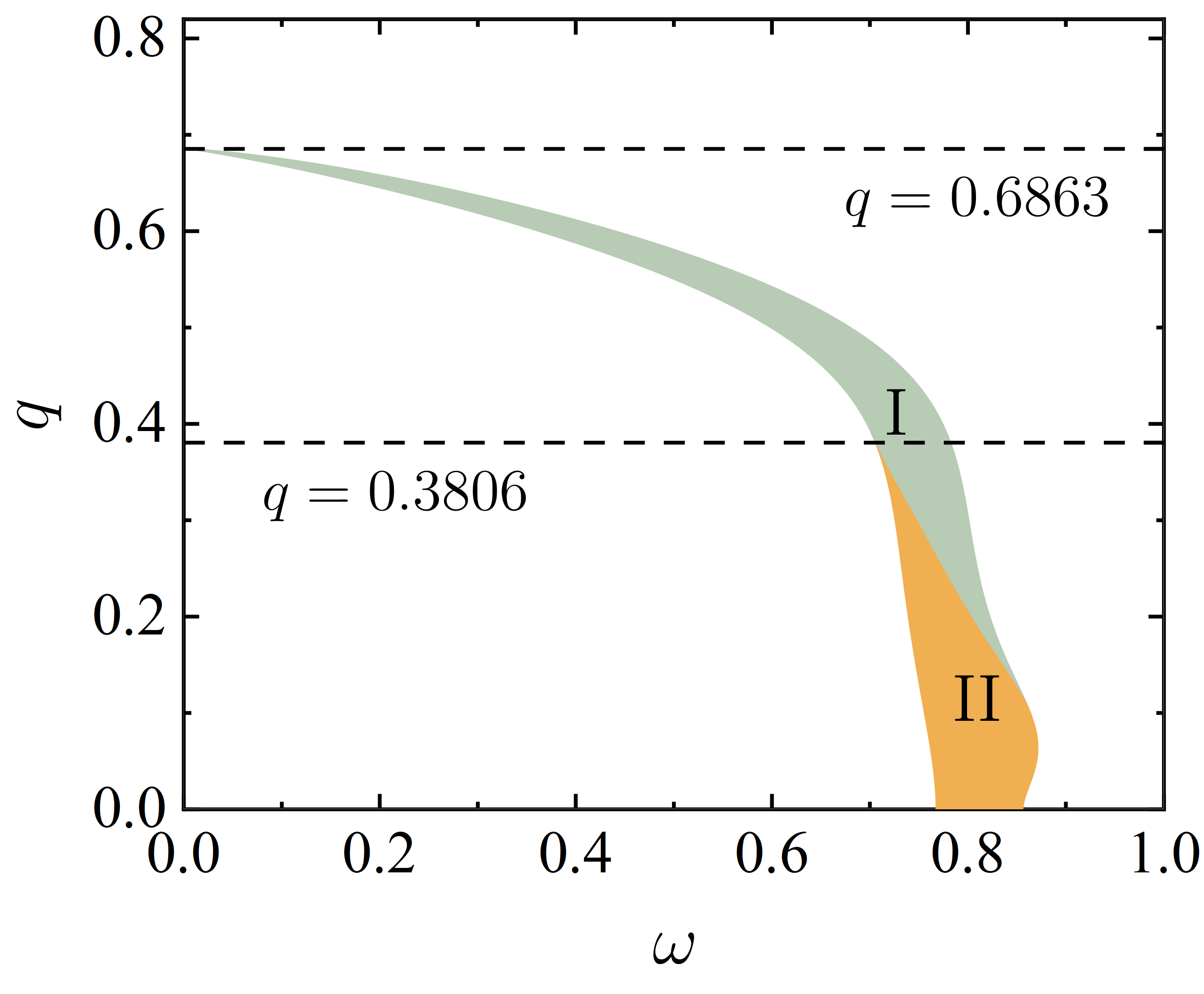}
   		} 		
		\caption{The domain of existence for solutions of ABGBSs in the $q-\omega$ plane for the first branch (left) and second branch (right). The regions are classified by which metric functions can exceed unity: neither in region I, only $1/g_{rr}$ in region II, and both $-g_{tt}$ and $1/g_{rr}$ in region III.}
    	    \label{fig:Domain}		
		\end{figure}

Fig.~\ref{fig:Domain} shows the domains of existence of ABGBS solutions for the first branch (left) and second branch (right) in the $q$–$\omega$ plane. In region I, neither of the metric functions $-g_{tt}$ and $1/g_{rr}$ exceeds unity. In region II, only $1/g_{rr}$ can exceed unity in some radial regions, whereas in region III, both functions can exceed unity in certain radial regions.

The figure shows that, for relatively large electric charges ($q\gtrsim0.4$), the maximum frequency decreases significantly as $q$ increases. In our numerical calculations, ABGBS solutions are found only for charges below a critical value $q_c\approx0.7942$. This numerical upper bound is close to, but slightly below, $\sqrt{s/\alpha_{\mathrm{s}}}\approx\sqrt{0.2/0.31709}\approx0.7942$ (by less than $10^{-5}$), the critical charge at which the pure regular ABG solution becomes an extremal black hole for fixed $s$\footnote{We emphasize that the maximum charge $q_c$ is obtained by computing solutions down to $\omega=0.001$. Considering the behavior shown in Fig.~\ref{fig:Domain}, which shows that the maximum frequency decreases noticeably as $q$ increases for relatively large charges, and the fact that we have not found ABGBS solutions with $\omega=0$, we conjecture that the maximum allowed charge of ABGBSs may approach $\sqrt{s/\alpha_{\mathrm{s}}}$ arbitrarily closely when $\omega\to0$.}. Within this charge range, both metric functions can exceed unity on the first branch when $q<0.4472$ and the frequency is relatively large. On the second branch, by contrast, only $1/g_{rr}$ can exceed unity, while $-g_{tt}$ remains at or below unity.

The upper bound on the charge of solutions with $1/g_{rr}>1$ near the center can also be inferred from the field equation for $n(r)$, Eq.~\eqref{eq:n}. Specifically, expanding this equation near $r=0$ yields
\begin{equation}
n(r)=1+\left[
\frac{1}{q^2}-\frac{1}{s}
-\frac{2}{3}\left(1+\frac{\omega^2}{\sigma_0^2}\right)\phi_0^2
\right]r^2+O(r^4),
\end{equation}
where $\phi_0=\phi(0)$ and $\sigma_0=\sigma(0)$. Thus, for $n(r)$ to exceed unity near the center, the coefficient of $r^2$ must be positive. Since the contribution from the scalar field to this
coefficient is nonpositive, a necessary condition is $q<\sqrt{s}\approx0.4472$. This is the same necessary condition as for the pure regular ABG solution (see Eq.~\eqref{eq:RABGF0}). 

Turning to the case of relatively large electric charges, we show the radial profiles of the metric functions $-g_{tt}=n\sigma^2$, $1/g_{rr}=n$, and $\sigma$, the potential $V$, the scalar field, and the energy density $\rho_{\phi}=-T^{(\rm S)t}{}_{t}$ in Fig.~\ref{fig:BFfrozen}. As the frequency decreases, the minimum values of the metric functions $-g_{tt}$, and $1/g_{rr}$ gradually decrease. In particular, for $\omega=0.001$, $g^{rr}$ becomes very close to zero at a critical position $\bar{r}_{\rm ch}$, while $-g_{tt}$ also becomes very close to zero inside $\bar{r}_{\rm ch}$. However, we do not find any ABGBS solutions in which either $-g_{tt}$ or $g^{rr}$ reaches zero. This behavior is characteristic of a frozen star. Moreover, since the behavior of these metric functions near the surface $\bar{r}=\bar{r}_{\rm ch}$ is similar to that near an event horizon, we refer to this as the ``critical horizon''.

	\begin{figure}[!htbp]
		\centering		
		\subfigure{  
			\includegraphics[width=7.2cm]{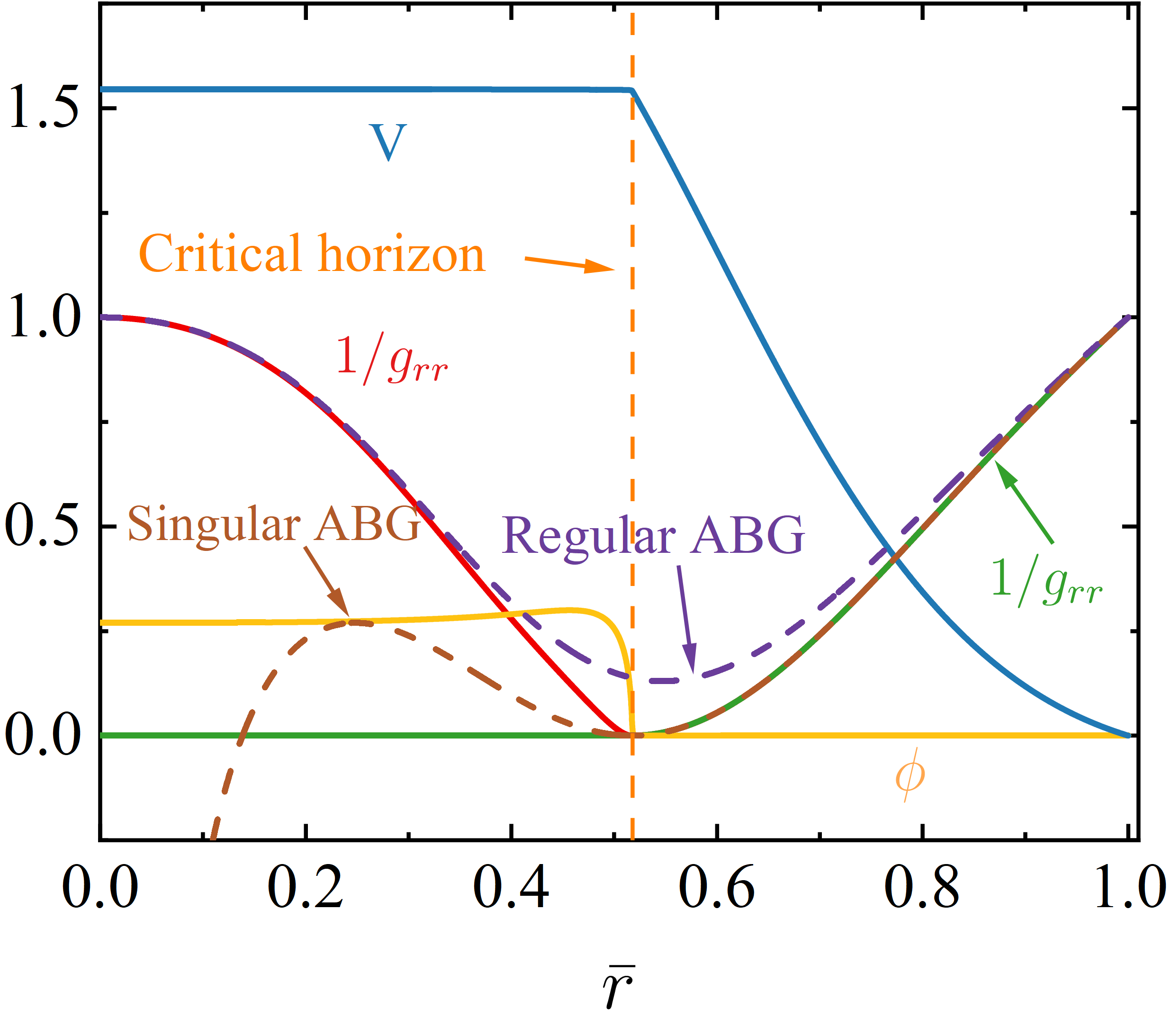}
		}		
		\caption{The field distributions of the ABGBS solutions with $\omega=0.001$ and the ABG solutions versus the compactified coordinate $\bar{r}$. The purple curves denote the pure regular ABG solution. The brown curves denote the pure singular extremal ABG BH with $C\approx-0.14813$. All solutions have $q=0.75$.}
    	    \label{fig:Extremal}		
		\end{figure}

For scalar and electromagnetic fields, as frequency decreases, the maximum values of the scalar field $\phi$ and its energy density $\rho_{\phi}$ gradually increase, while the maximum value of the electric potential $V$ decreases. As the frequency approaches zero, the profiles of the scalar field and its energy density develop steep, wall-like transitions near the critical position $\bar{r}_{\rm ch}$. Beyond $\bar{r}_{\rm ch}$, the scalar field nearly vanishes. Meanwhile, the electric potential $V$ remains nearly constant inside $\bar{r}_{\rm ch}$.

 Fig.~\ref{fig:Extremal} shows the radial profiles of the field function for the ABGBS solutions at $\omega=0.001$, together with the metric for the regular ABG solution and the singular extremal ABG BH solution, for the same value $q=0.75$. It can be seen that, after introducing the scalar field, the minimum values of the metric functions for the resulting ABGBSs are lower than those for the regular ABG solution. Moreover, outside the event horizon of the extremal black hole, the metric functions $-g_{tt}$ and $1/g_{rr}$ of the ABGBS solutions coincide with those of the extremal black hole. Notably, the masses of the singular extremal ABG black hole and the ABGBS are nearly identical, with both approximately equal to $M=1.12875$. These results suggest that these solutions can be regarded as an extremal black hole from the vantage point of an observer at infinity.

\subsection{Timelike geodesics}\label{sec:pemodel}

Our results show that ABGBSs exhibit a variety of properties that differ from those of ordinary mini boson stars. In particular, the metric properties of frozen ABGBSs are very close to those of extremal black holes. They may therefore exhibit observational features that distinguish them from ordinary boson stars. In this section, we mainly study unbound timelike trajectories in frozen ABGBS spacetimes to provide an initial exploration of the properties of these BSs. Given the distinctive behavior of the metric functions near the critical horizon, we focus on trajectories that cross it.

Consider a massive test particle moving around the ABGBSs. The Lagrangian of the particle is 
\begin{equation}
\mathcal{L}=\frac{1}{2}g_{\mu\nu}\dot{x}^{\mu}\dot{x}^{\nu},
\label{eq:PL}
\end{equation}
where the overdot denotes differentiation with respect to the proper time $\tau$. Given the spherical symmetry of the spacetime, without loss of generality, we focus on equatorial geodesics with $\theta=\pi/2$. Then one can obtain the equation from the Lagrangian \eqref{eq:PL} as
\begin{equation}
g_{tt} \dot{t}^2+g_{rr} \dot{r}^2+g_{\varphi\varphi} \dot{\varphi}^2=-1.
\label{eq:Pds}
\end{equation}

Since the spherically symmetric system possesses two Killing vectors $\partial_{t}$ and $\partial_{\varphi}$, one can define the energy $E=-g_{tt}\dot{t}$ and the angular momentum $L=r^2 \dot{\varphi}$ of the particle as two conserved quantities. Exploiting the two conserved quantities, the specific energy and the specific angular momentum, we have 
\begin{equation}
\dot{r}^2=\frac{E^2}{\sigma(r)^2}-n(r)\left(1+\frac{L^2}{r^2}\right).
\label{eq:dotr}
\end{equation}
The effective potential is an easy way to determine the radial motion of a particle, which can be obtained by setting $\dot{r}^2=0$, yielding
\begin{equation}
V_{\mathrm{eff}}=\sigma(r)\sqrt{n(r)\left(1+\frac{L^2}{r^2}\right)}.
\end{equation}

\begin{figure}[!htbp]
		\centering		
		\subfigure{  
			\includegraphics[width=7.2cm]{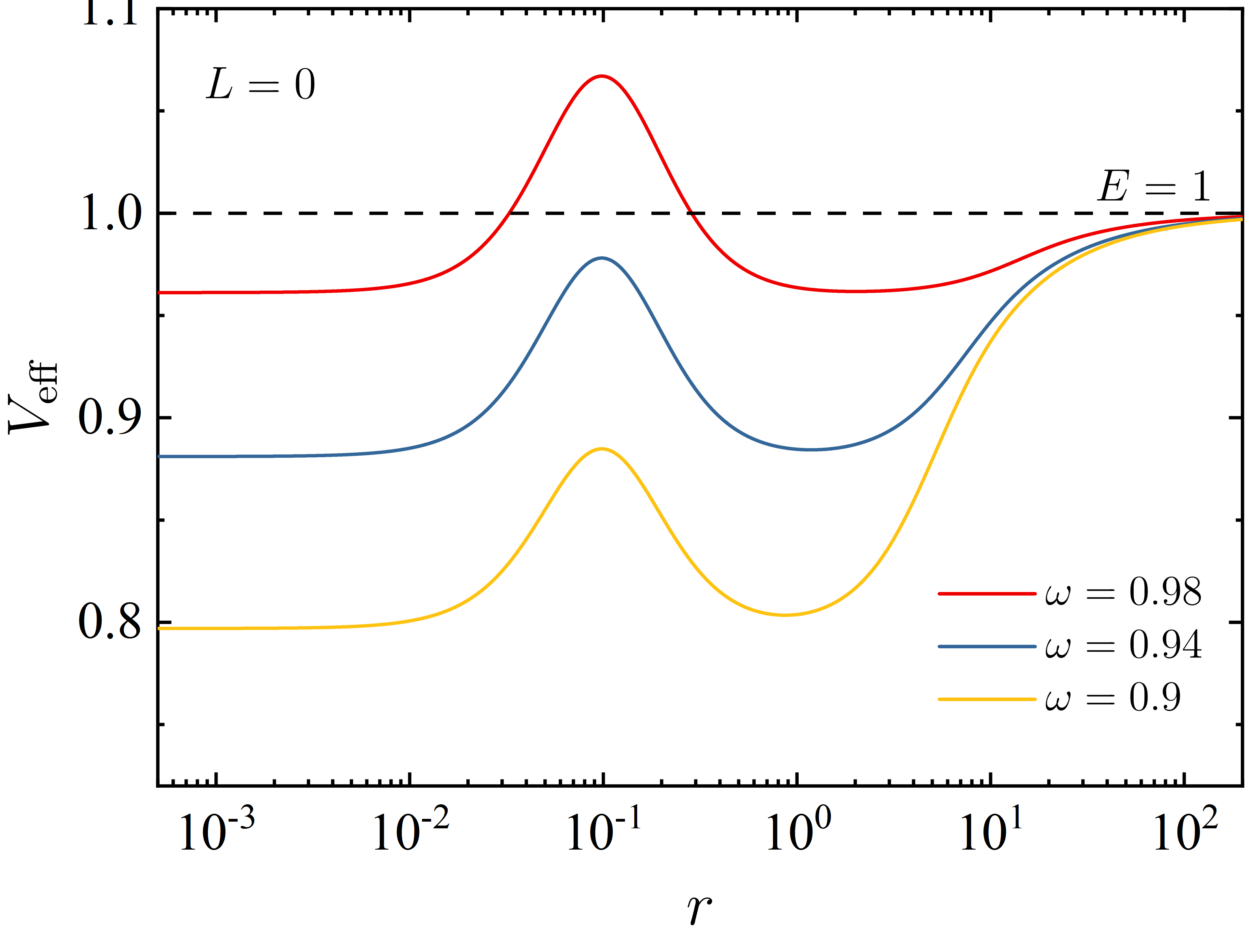}
		}		
		\caption{Effective potential with $L=0$ as a function of radius $r$ for ABGBSs with $q=0.1$.}
    	    \label{fig:PbNxFun}		
		\end{figure}

 For radial trajectories passing through the center, the angular momentum vanishes, $L=0$, and the effective potential reduces to
\begin{equation}
V_{\mathrm{eff}}(r)=\sigma(r)\sqrt{n(r)}
=\sqrt{-g_{tt}(r)}.
\end{equation}
Therefore, as shown in Fig.~\ref{fig:PbNxFun}, for BSs exhibiting gravitational blueshift,
 $V_{\mathrm{eff}}>1$ in the region where $-g_{tt}>1$.
For the ordinary mini boson star considered here, by contrast, the radial effective potential does not exceed unity. This implies that, compared with the ordinary boson star, a particle coming from infinity requires more energy to cross the potential barrier and reach the center of a blueshifted ABGBS.

For trajectories that do not pass through the center, according to $E=-g_{tt}\dot{t}$ and Eq.~\eqref{eq:dotr}, the radial velocity of the test particle is given by
\begin{equation}
\left(\frac{\mathrm{d}r}{\mathrm{d}t}\right)^2=\left(\frac{\mathrm{d}r}{\mathrm{d}\tau}\frac{\mathrm{d}\tau}{\mathrm{d}t}\right)^2=\frac{n^2\sigma^2}{E^2}
\left[
E^2-n\sigma^2
\left(1+\frac{L^2}{r^2}\right)
\right].
\end{equation}
The angular velocity is obtained from $L=r^2\dot{\varphi}$ and $E=-g_{tt}\dot{t}$, i.e.,
\begin{equation}
\frac{\mathrm{d}\varphi}{\mathrm{d}t}
=\frac{\dot\varphi}{\dot t}
=\frac{n(r)\sigma^2(r)L}{E r^2}.
\label{eq:phi}
\end{equation}
By integrating the radial and angular velocities, we can plot the trajectories of the test particles.
	\begin{figure}[!htbp]
\begin{center}
\subfigure{
\includegraphics[width=3.8cm]{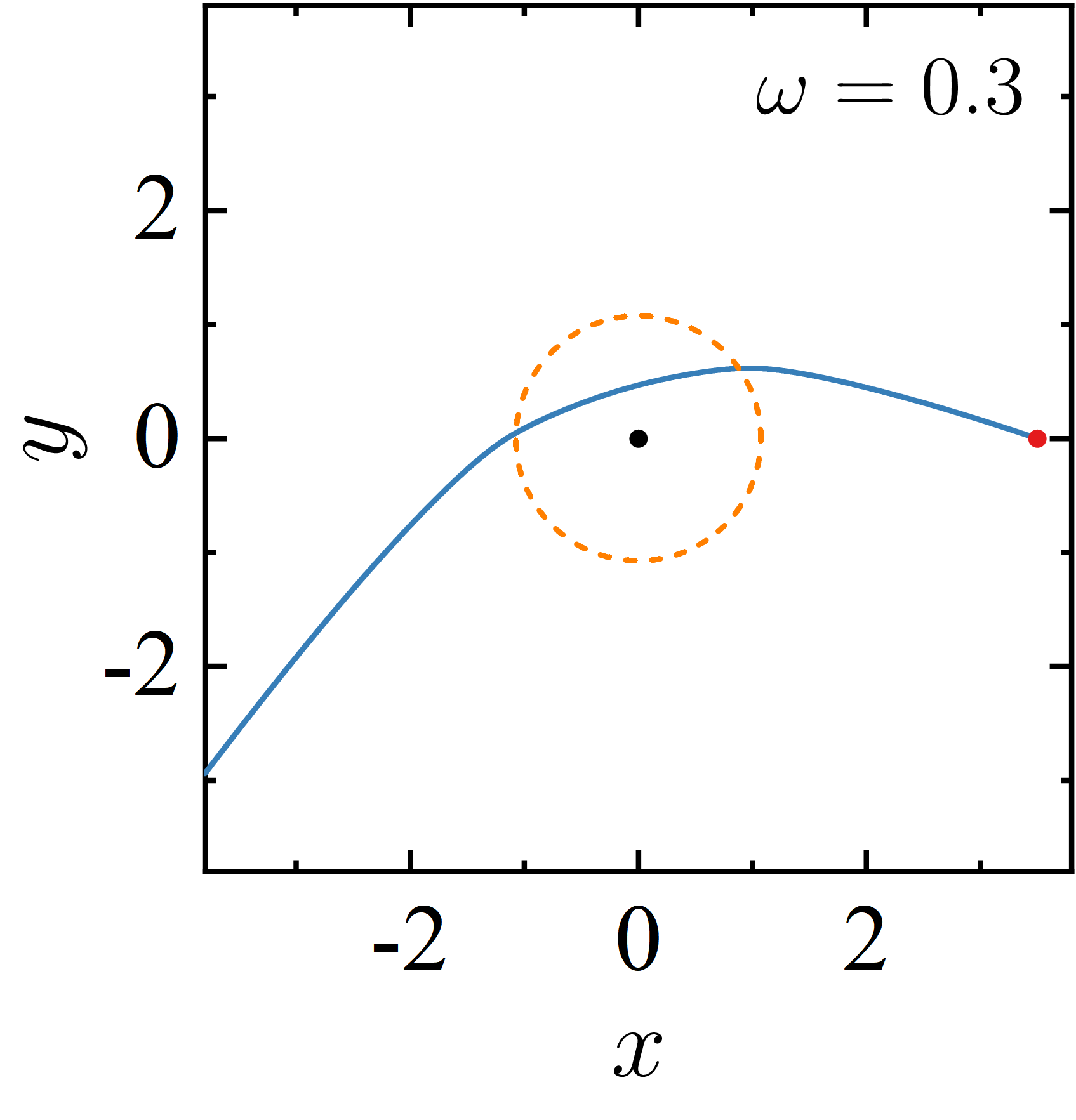}}
\subfigure{
\includegraphics[width=3.8cm]{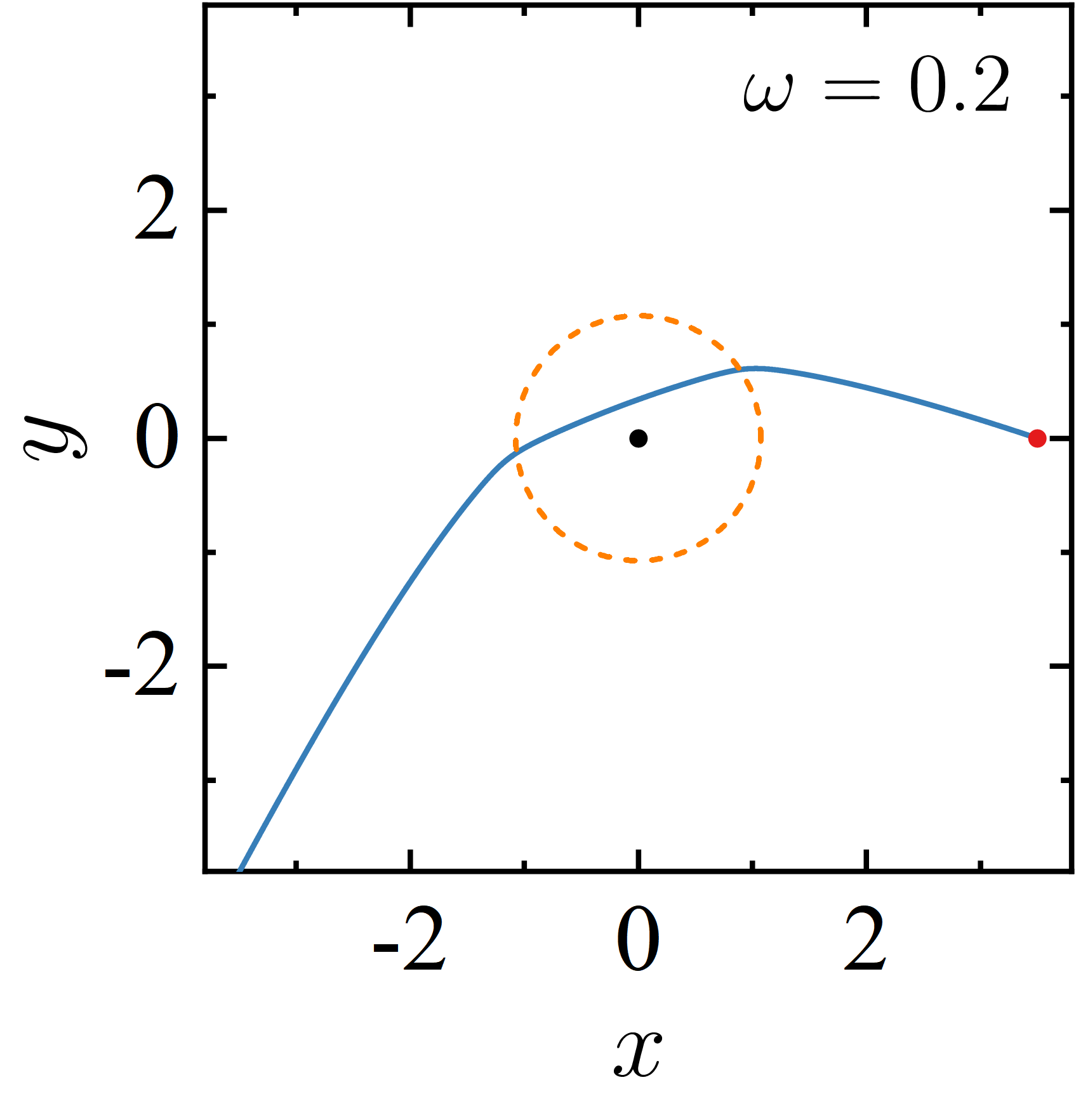}}
\subfigure{
\includegraphics[width=3.8cm]{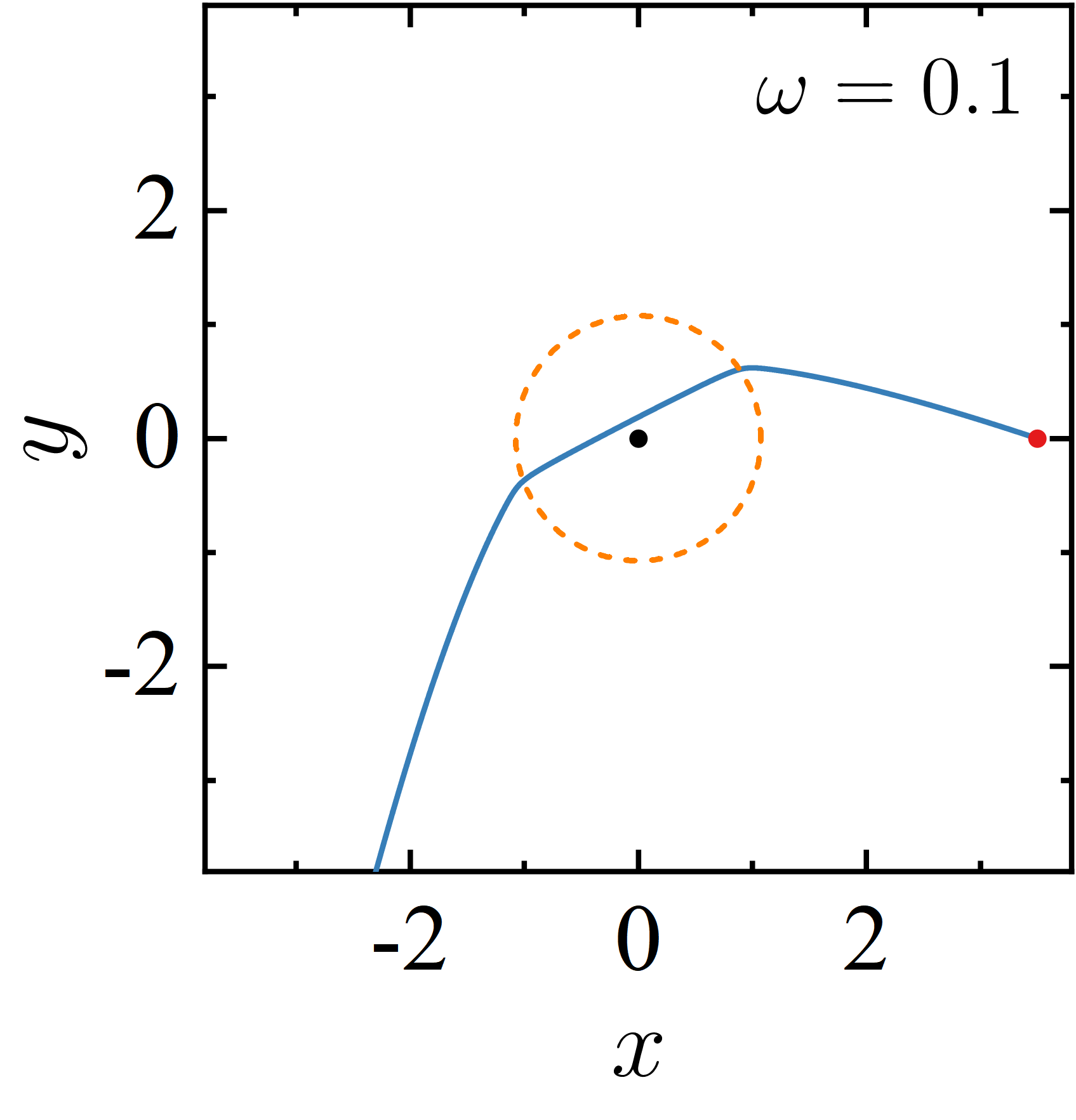}}
\subfigure{
\includegraphics[width=3.8cm]{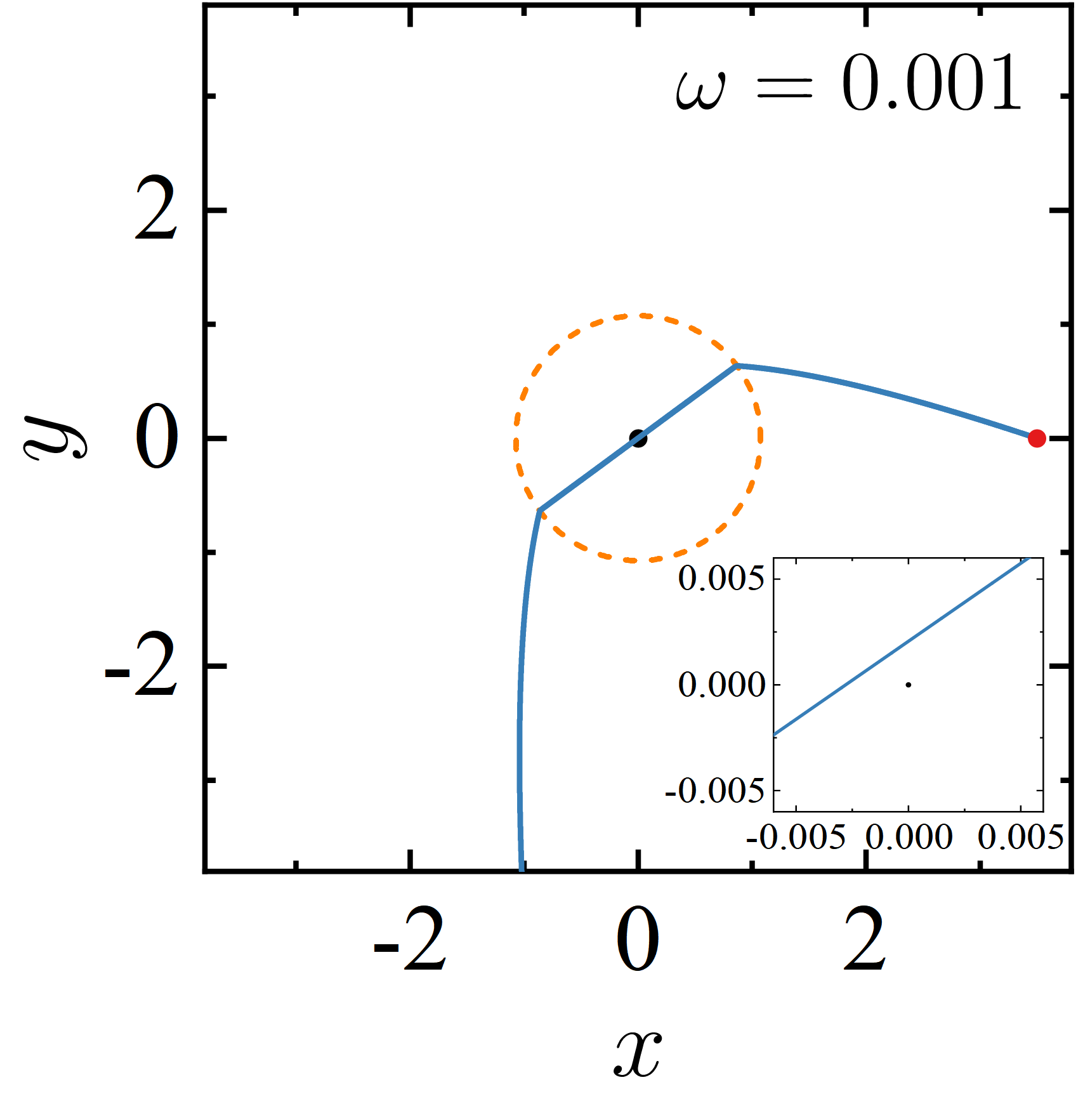}}	
\end{center}
		\caption{Timelike orbits around ABGBSs with $q=0.75$ for several values of the scalar frequency $\omega$, computed for
        ($E=1.1$, $L=1$), where $x=r\cos\varphi$ and $y=r\sin\varphi$.}
		\label{fig:Orbit}		
		\end{figure}

As an example, Fig.~\ref{fig:Orbit} shows timelike trajectories in ABGBS spacetimes with $q=0.75$ for different values of the frequency, where the orange dashed line represents the critical horizon $r_{\mathrm{ch}}$ of ABGBSs with $\omega=0.001$, and the initial position of the timelike particle is $(x,y)=(3.5,0)$ (red dot). It can be seen that, for a larger frequency $\omega=0.3$, the trajectory of the test particle is relatively weakly deflected. As $\omega$ decreases, the trajectory is deflected more strongly, and the test particle can pass closer to the center. For $\omega=0.001$, corresponding to a frozen ABGBS, the trajectory undergoes a large deflection near the critical horizon. Inside the critical horizon, it appears nearly straight and passes very close to the center (black dot).

\begin{figure}[htbp]
\begin{center}
\subfigure{
\includegraphics[width=3.8cm]{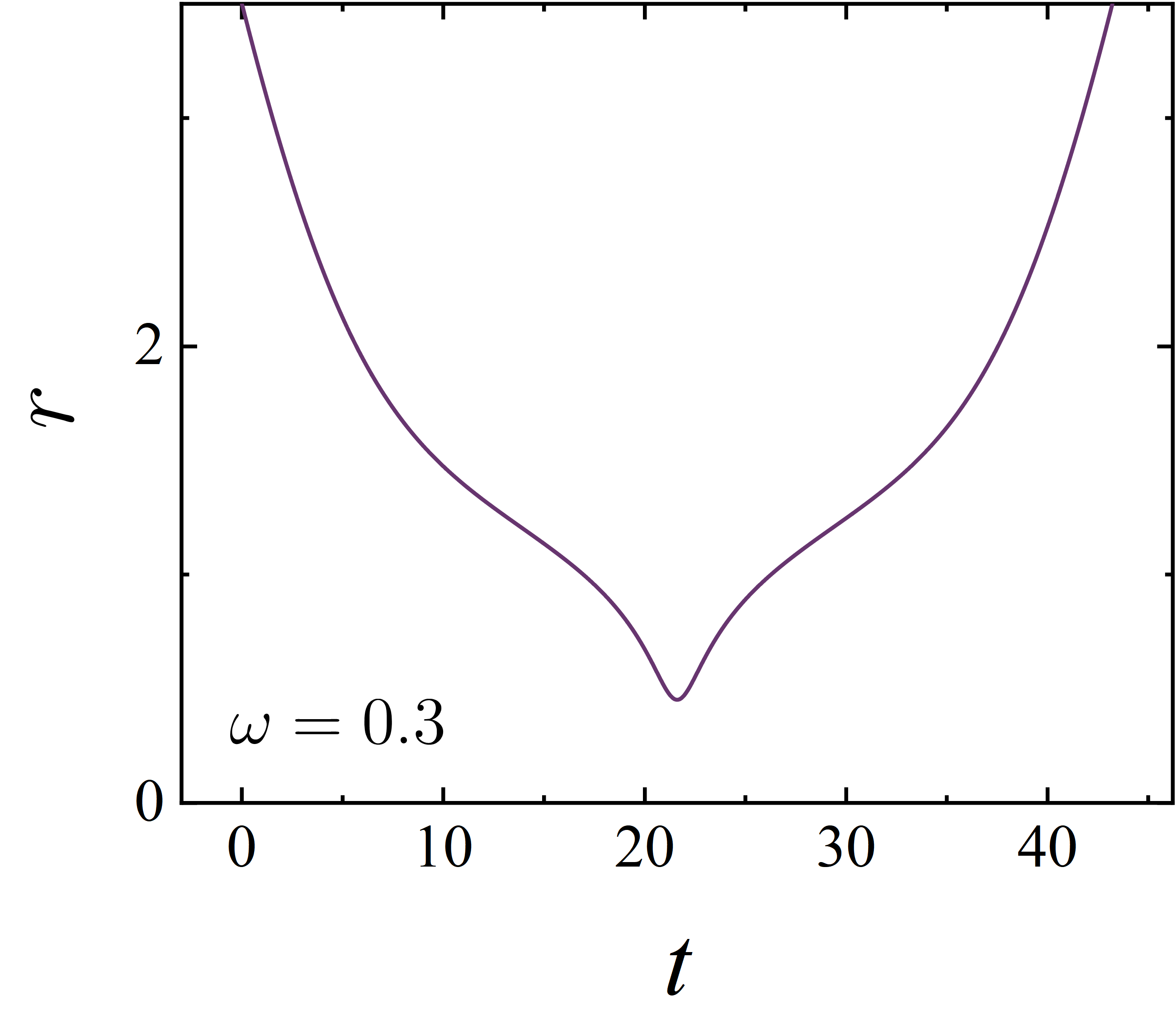}}
\subfigure{
\includegraphics[width=3.8cm]{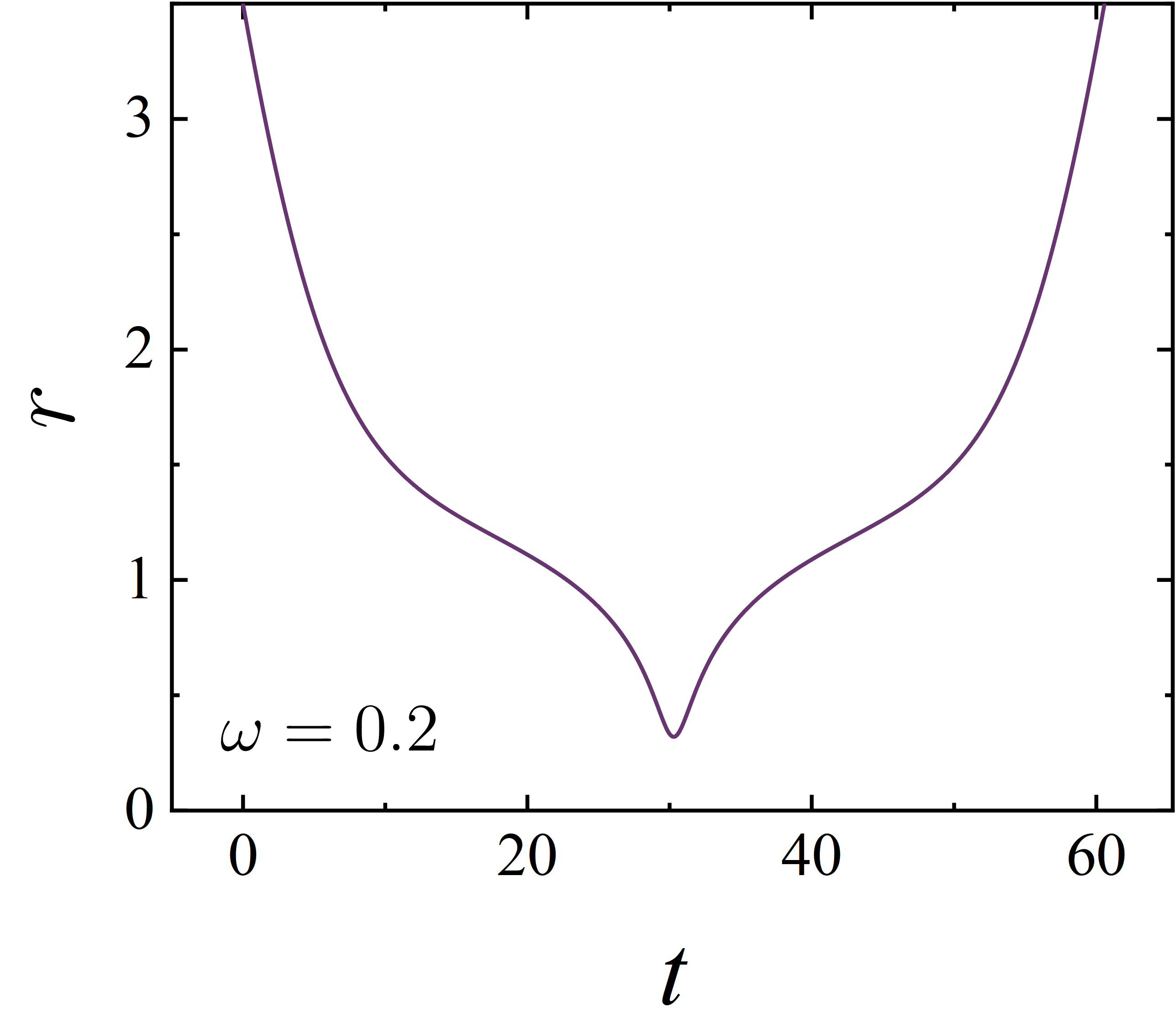}}
\subfigure{
\includegraphics[width=3.8cm]{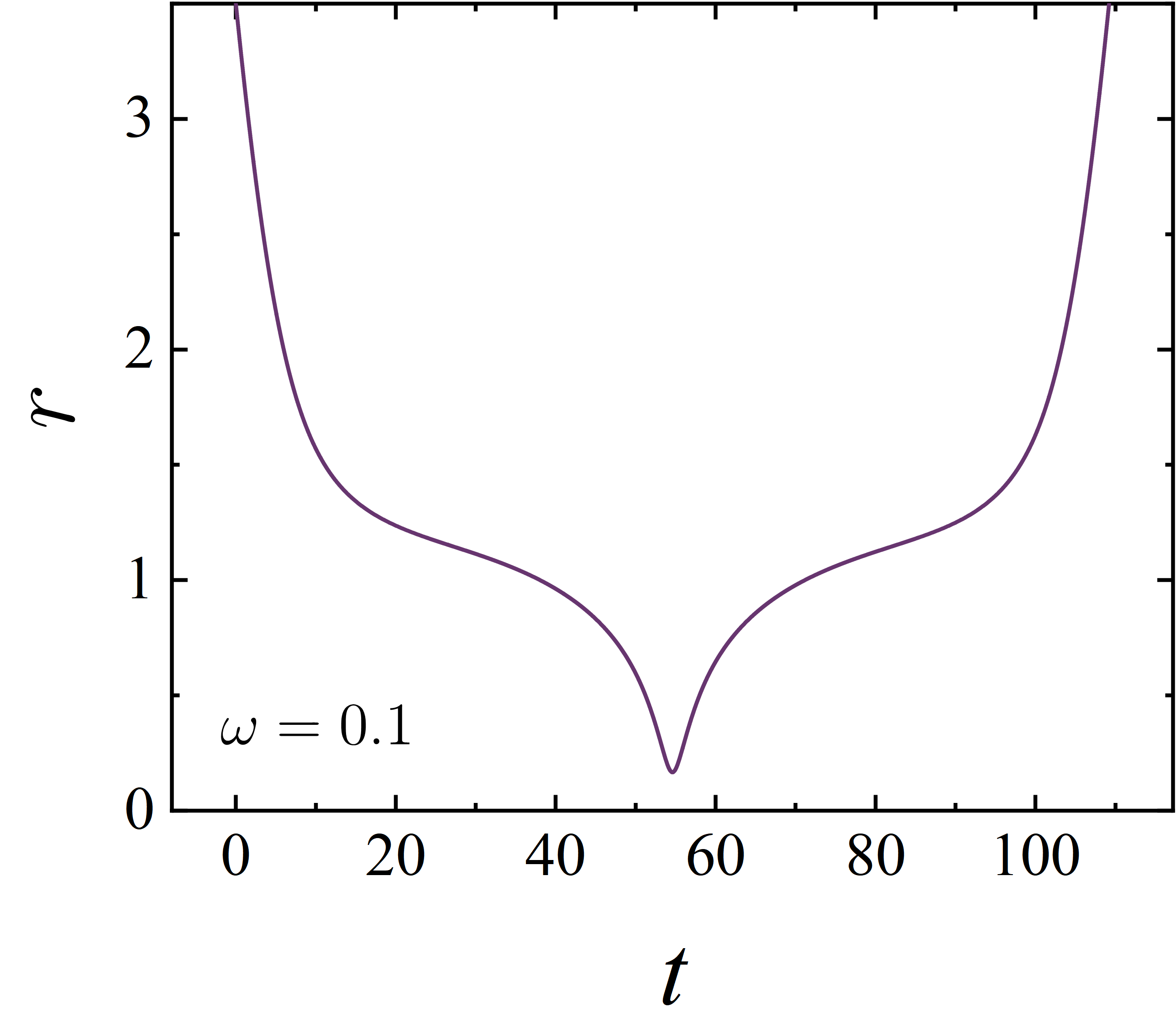}}
\subfigure{
\includegraphics[width=3.8cm]{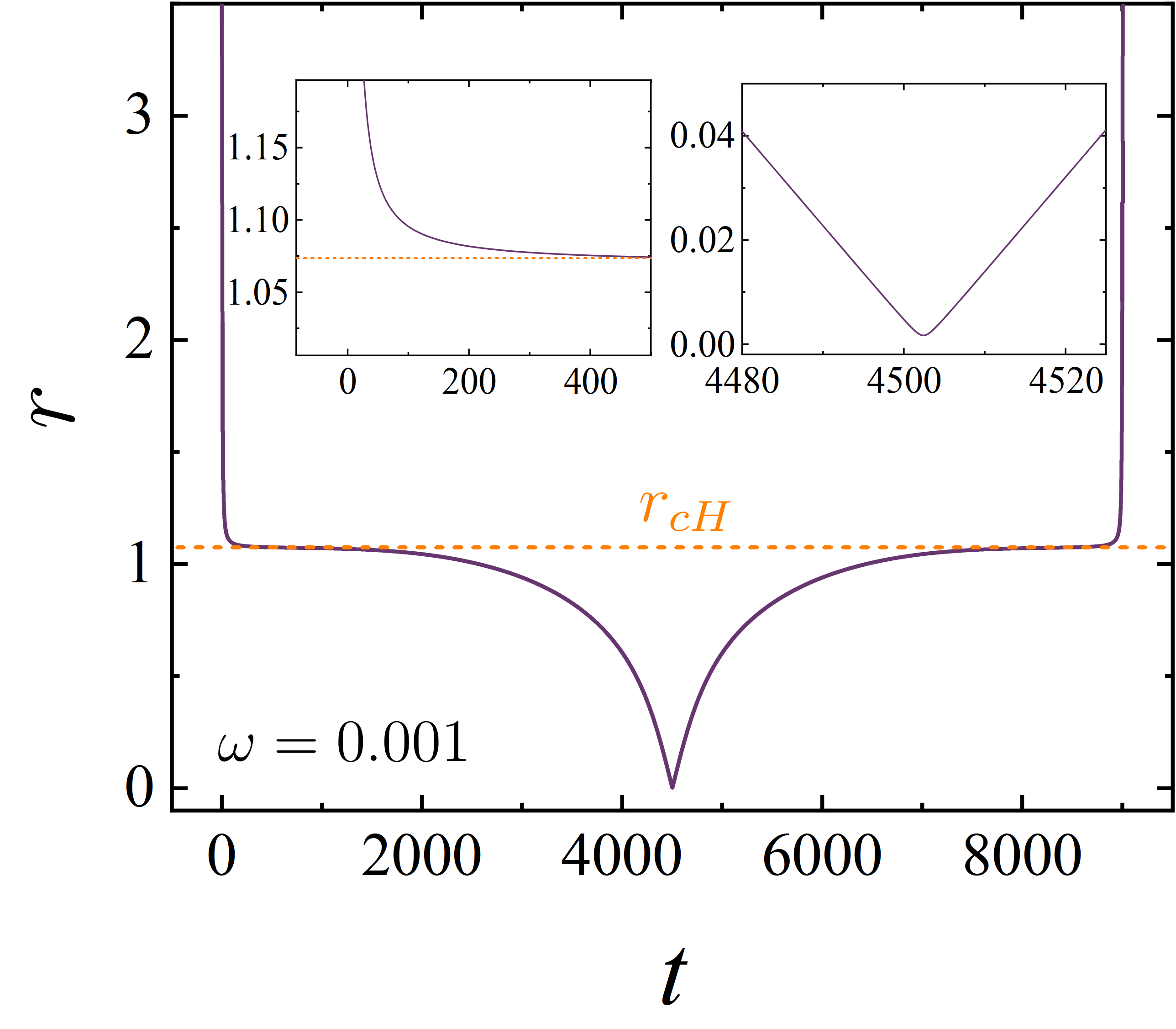}}	
\end{center}
    \caption{The relationship between radial distance $r$ and time $t$ of the timelike geodesic around ABGBSs with $q=0.75$, $E=1.1$, and $L=1$.}
    \label{fig:rt}
\end{figure}

The corresponding evolution of the radial coordinate of the test particles is shown in Fig.~\ref{fig:rt}. As $\omega$ approaches zero, the coordinate time spent by the particle within the same radial distance increases. For the frozen solution with $\omega=0.001$, this time is roughly two orders of magnitude longer than for the solution with $\omega=0.3$, and the particle spends a very long coordinate time near the critical horizon.

\section{CONCLUSION}\label{sec: conclusion}

In this work, we investigated the model of Einstein-Klein-Gordon theory coupled to the nonlinear electrodynamics proposed by E.~Ayon-Beato and A.~Garcia, and found ABGBS solutions with electric charge. Our results reveal that ABGBSs exist only for electric charges below a critical value $q_c$, and that the charge plays a crucial role in determining their properties. Compared with Bardeen BSs and Hayward BSs with magnetic charge~\cite{Wang:2023tdz,Yue:2023sep}, these ABGBSs with electric charge exhibit a richer variety of properties.

As the electric charge $q$ increases, the maximum scalar-field frequency gradually decreases. Unlike ordinary boson stars, the solutions at the maximum frequency no longer approach a dilute limit but instead approach a pure regular ABG solution. In the first branch solutions, for small electric charges ($q<0.4472$) and relatively large frequencies, the metric functions $-g_{tt}$ and $1/g_{rr}$ can exceed unity in certain regions, corresponding to gravitational blueshift and a negative mass function, respectively. Compared with ordinary boson stars, particles require more energy to pass through the center of these boson stars.

For larger charges ($q>0.6863$), the multiple branch solution vanishes, and frozen ABGBS solutions can be obtained when the frequency of scalar field is close to zero. In these solutions, the scalar field is concentrated near the critical horizon and decays rapidly outside it, while its energy density exhibits a steep, wall-like distribution near the inner side of the critical horizon. Meanwhile, the electromagnetic potential remains nearly constant inside the critical horizon. Outside the critical horizon, the spacetime of a frozen ABGBS resembles that of an extremal black hole with nearly the same mass.

Although the spacetime properties of these frozen solutions resemble those of black holes, they remain boson-star solutions, and their timelike geodesics are complete. In these spacetimes, timelike particle trajectories are strongly deflected near the critical horizon, and particles take a longer coordinate time to move inside it. Similarly, previous studies~\cite{Kagramanova:2007tt,Huang:2025css} have also observed these characteristics for null geodesics in analogous frozen spacetimes, reflecting the distinctive orbital properties of these spacetimes.

Finally, it is worth noting that our model can also be viewed as introducing a scalar field into the regular ABG spacetime. The pure regular ABG solution describes a BH when the electric charge is equal to or greater than $\sqrt{s/\alpha_{\mathrm{s}}}\approx0.7942$. This critical value is close to, but slightly greater than, the maximum charge $q_c$ of the ABGBS solutions. Therefore, our results appear to suggest that the presence of a nontrivial scalar field limits the increase of charge, and prevents the formation of a regular black hole, similar to what has been found for magnetically charged Bardeen and Hayward boson stars~\cite{Wang:2023tdz,Yue:2023sep}.
From this perspective, these studies and our results together appear to suggest a possible no-hair theorem under the requirement of regularity throughout spacetime.

Several extensions of our study merit further exploration. First, the complex scalar field considered in this work is ungauged. An interesting extension would be to introduce a gauge coupling and investigate how it affects the properties of ABGBSs. Second, in the magnetic case, previous studies have examined thin-disk images of Hayward boson stars~\cite{Zhao:2025yhy} and Bardeen boson stars~\cite{Yang:2025usj}, finding that their images can exhibit features similar to those of BHs. Since frozen ABGBSs share metric features with BHs, extending such analyses to electrically charged ABGBSs could provide further insights into their observational signatures. Lastly, the dynamical stability of these solutions remains an important subject for future investigation.

\section*{ACKNOWLEDGEMENTS}
	This work is supported by the National Key Research and Development Program of China (Grant Nos. 2022YFC2204101 and 2020YFC2201503), the National Natural Science Foundation of China (Grant Nos. 12275110, 12675069 and  12247101) and the Fundamental Research Funds for the Central Universities (Grants No. lzujbky-2024-eyt02).

\end{document}